\documentclass[fleqn,usenatbib]{mnras}

\usepackage{graphicx}	
\usepackage{amsmath}	

\usepackage{amsmath}
\usepackage{hyperref}
\usepackage{datetime}
\usepackage{appendix}
\usepackage{comment}
\usepackage{chngcntr}
\usepackage{soul}
\usepackage{bm}
\usepackage{xcolor}
\usepackage{xspace}

\def\redmapper{\rm redMaPPer\xspace}
\def\msun{h^{-1}{M_\odot}}
\def\mpc{h^{-1}{\rm Mpc}}
\def\gpc{h^{-1}{\rm Gpc}}

\newcommand{\bx}{{\mathbf{x}}}
\newcommand{\bk}{{\mathbf{k}}}
\newcommand{\br}{{\mathbf{r}}}

\newcommand{\hbk}{\hat{\mathbf{k}}}
\newcommand{\hbn}{\hat{\mathbf{n}}}
\newcommand{\hbr}{\hat{\mathbf{r}}}
\newcommand{\avrg}[1]{\left\langle #1 \right\rangle}
\newcommand{\rmD}{\mathrm{D}}
\newcommand{\rmg}{\mathrm{g}}
\newcommand{\rmd}{\mathrm{d}}

\newcommand{\rmm}{\mathrm{m}}
\newcommand{\delg}{\delta_{\rmg}}

\title[IA of Clusters]{The Intrinsic Alignment of Galaxy Clusters and Impact of Projection Effects}

\author[J. Shi et al.]{
Jingjing Shi$^{1,2}$ \thanks{E-mail:jingjing.shi@ipmu.jp}, Tomomi Sunayama$^{3,4}$, Toshiki Kurita$^{1}$, Masahiro Takada$^{1,2}$, Sunao Sugiyama${}^{1,2}$, 
\newauthor
Rachel Mandelbaum${}^{5}$, Hironao Miyatake${}^{3}$, Surhud More${}^{6,1}$, Takahiro Nishimichi,${}^{8,7,1}$, Harry Johnston$^{9}$
\\
\\
$^{1}$Kavli Institute for the Physics and Mathematics of the Universe (WPI), 
The University of Tokyo Institutes for Advanced Study (UTIAS), \\
The University of Tokyo, 5-1-5 Kashiwanoha, Kashiwa-shi, Chiba, 277-8583, Japan \\
$^{2}$Center for Data-Driven Discovery (CD3),  Kavli IPMU (WPI), UTIAS, The University of Tokyo, Kashiwa, Chiba 277-8583, Japan \\
$^{3}$Kobayashi-Maskawa Institute for the Origin of Particles and the Universe (KMI), Nagoya University, Nagoya 464-8602, Japan \\
$^{4}$Department of Astronomy and Steward Observatory, University of Arizona, 933 N Cherry Ave, Tucson, AZ, 85719, USA\\
$^{5}$McWilliams Center for Cosmology, Department of Physics, Carnegie Mellon University, Pittsburgh, PA 15213, USA \\
$^{6}$The Inter-University Centre for Astronomy and Astrophysics,
Post bag 4, Ganeshkhind, Pune 411007, India \\
$^{7}$Center for Gravitational Physics and Quantum Information, Yukawa Institute for Theoretical Physics, Kyoto University,\\
Kyoto 606-8502, Japan \\
$^{8}$Department of Astrophysics and Atmospheric Sciences, Faculty of Science, Kyoto Sangyo University, Kyoto 603-8555, Japan\\
$^{9}$Institute for Theoretical Physics, Utrecht University, Princetonplein 5, 3584 CE Utrecht, The Netherlands
}

\date{Accepted XXX. Received YYY; in original form ZZZ}

\pubyear{2023}

\begin{document}
\label{firstpage}
\pagerange{\pageref{firstpage}--\pageref{lastpage}}
\maketitle

\begin{abstract}
Galaxy clusters, being the most massive objects in the Universe, exhibit the strongest alignment with the large-scale structure. However, mis-identification of members due to projection effects from the large scale structure can occur. We studied the impact of projection effects on the measurement of the intrinsic alignment of galaxy clusters, using galaxy cluster mock catalogs. 
Our findings showed that projection effects result in a decrease of the large scale intrinsic alignment signal of the cluster and produce a bump at $r_p\sim 1\mpc$, most likely due to interlopers and missed member galaxies. This decrease in signal explains the observed similar alignment strength between bright central galaxies and clusters in the SDSS $\redmapper$ cluster catalog. The projection effect and cluster intrinsic alignment signal are coupled, with clusters having lower fractions of missing members or having higher fraction of interlopers exhibiting higher alignment signals in their projected shapes. We aim to use these findings to determine the impact of projection effects on galaxy cluster cosmology in future studies.
\end{abstract}


\section{Introduction}

Galaxy clusters are a major probe of dark energy \citep{2013PhR...530...87W}. Their abundance and time evolution are sensitive to the growth of structure in the Universe, since they form from rare highest peaks of the initial density field. Cluster cosmology is a major science of many surveys, including Hyper Suprime-Cam (HSC) survey \footnote{https://hsc.mtk.nao.ac.jp/ssp/}, the Dark Energy Survey (DES) \footnote{https://www.darkenergysurvey.org/}, the Kilo Degree Survey (KiDS) \footnote{https://kids.strw.leidenuniv.nl/}, the Rubin Observatory Legacy Survey of Space and Time (LSST) \footnote{https://www.lsst.org/}, Euclid \footnote{https://www.euclid-ec.org/}, and the Nancy Grace Roman Telescope \footnote{https://roman.gsfc.nasa.gov/}.    

Cluster shapes are triaxial, originating from the anisotropic matter field and accretion. As a result, cluster shapes are expected to align with the matter field, i.e.\ intrinsic alignment (IA) (see review papers by \citealt{2015SSRv..193....1J, 2015PhR...558....1T, 2015SSRv..193..139K, 2015SSRv..193...67K}). IA are distinct from the alignments of galaxy shapes that originate from gravitational lensing by foreground attractors. The IA signal has been observed for massive red galaxies \citep{2009ApJ...694..214O,2015MNRAS.450.2195S}, but no clear detection has been claimed for blue galaxies \citep{2011MNRAS.410..844M, 2020ApJ...904..135Y}. 
The alignment of galaxy clusters have also been detected \citep{2012MNRAS.423..856S}. \cite{2017MNRAS.468.4502V} studied the cluster shape - density correlation using $\redmapper$ clusters from Sloan Digital Sky Survey-Data Release 8 (SDSS DR8), finding a higher IA amplitude of galaxy clusters than luminous red galaxies (LRGs). As clusters are the most massive bound structures, studies on cluster shapes offer the unique opportunity to yield insight into dark matter halo shapes \citep{2009ApJ...695.1446E, 2010MNRAS.405.2215O,2018MNRAS.475.2421S,2022MNRAS.517.4827G}.

However, the IA amplitude of galaxy clusters are found to be lower than predictions from numerical $N$-body simulations based on $\Lambda$ cold dark matter ($\Lambda$CDM) cosmology. \cite{2012MNRAS.423..856S} discussed various systematic observational uncertainties that may have caused this discrepancy, including photometric redshift error, cluster centroiding error, uncertainty in cluster shape estimation using a limited subsample of galaxy members, and inclusion of spherical clusters. However, one of the major systematics for optically identified clusters, the so-called ``projection effect", has not been properly discussed for measurement of IA for galaxy clusters. 

Projection effects refer to the fact that interloper galaxies along the line-of-sight (LOS) are mistakenly identified as members of galaxy clusters \citep{1997MNRAS.287..817V,2007MNRAS.382.1738C}. This is a major systematics for optical clusters whose mass proxy is a number of member galaxies (called richness). It can also boost cluster lensing and clustering signals on large scales, since clusters with a filamentary structure aligned with the LOS direction are preferentially identified by optical cluster finders, which typically detect clusters using red galaxy overdensities in photometric catalogs \citep{2018MNRAS.477.2141O,2020MNRAS.496.4468S,2022arXiv220503233S}.  
To obtain unbiased cosmological constraints using galaxy clusters, the projection effect has to be corrected or modelled accurately \citep{2021PhRvL.126n1301T, 2023MNRAS.518.5171P, 2019MNRAS.482..490C}.  

In this work, we will study the impact of projection effects on measurements of cluster IA with the aim to understand the measured IA of the most massive objects. We also search for new perspectives on projection effects and possible ways to mitigate the impacts on cluster observables. We found that the projection effects can largely explain the lower signal of observed cluster IA compared to that of simulated dark matter halos. 

The structure of the paper is organized as follows. In 
Section~\ref{sec: method}, we introduce our methodology for measuring the correlation function and modeling the signals. In Section~\ref{sec: data}, we introduce the observational data and mock simulation used in this paper. The results on measured IA in observation and mocks --- including the impact of projection effects --- are presented in Section~\ref{sec: results} and Section~\ref{sec: discuss}. In Section~\ref{sec: sum}, we summarize our results. 

\section{Methodology - Linear Alignment Model}
\label{sec: method}

In this section we briefly describe the leading theory of IA, i.e. the linear alignment model \citep{2001MNRAS.320L...7C,2004PhRvD..70f3526H}, and then define the model to use for the comparison with the IA measurements of the $\redmapper$ clusters. 

The linear alignment model predicts that the intrinsic shape of dark matter halos, and galaxy clusters in this paper, is determined by the gravitational tidal field at the time of formation of the halo or galaxy cluster. That is, the intrinsic ``shear'', which characterizes the shape of galaxy cluster, is given as
\begin{align}
(\gamma_1,\gamma_2)=-\frac{C_1}{4\pi G}(\partial_x^2-\partial_y^2,\partial_x\partial_y)\Phi_p,
\label{eq:LA}
\end{align}
where $\Phi_p$ is the primordial gravitational field and $C_1$ is a constant. Here
we take the ($x,y$) coordinates to be on the 2D plane perpendicular to the LOS direction.  
Throughout this paper, we employ a distant observer approximation, and in the above equation we take the LOS direction to be along the $z$-axis direction.

In this paper, we consider the cross-correlation between the IA shear of galaxy clusters and the galaxy density field.
For  the latter, we will use the spectroscopic sample of galaxies in the measurement.
We can define the coordinate-independent cross-correlation function as 
\begin{align}
\xi_{\rm g+}({\bf r})&\equiv {\avrg{\gamma_+({\bf x}; {\bf x}')\delta_{g}({\bf x}')}},
\end{align}
with $\gamma_+$ being defined as
\begin{align}
\gamma_+({\bf x};{\bf x}')\equiv 
\Re\left[\left(\gamma_1({\bf x})+ i\gamma_2({\bf x})\right)e^{-2i\phi_{\bf r}}\right].
\label{eq:shear+_def}
\end{align}
Here $\Re$ denotes a notation to take the real part of the cluster shear, ${\bf r}\equiv {\bf x}-{\bf x}'$, 
and $\phi_{{\bf r}}$ is the angle measured from the first coordinate axis to the projected separation vector ${\bf r}_p$ on the sky plane perpendicular to the LOS direction. 
Since we can measure only the projected shape of each cluster and the positions of clusters and galaxies are modulated by redshift-space distortion (RSD) \citep{Kaiser1987:RSD}, the 3D cross-correlation function is generally given as 
a function of the 3D separation vector ${\bf r}=(r_\parallel,{\bf r}_p)$, where $r_\parallel$ is the component parallel to the LOS direction and ${\bf r}_p$ is the 2D separation vector perpendicular to the LOS.

Following the formulation in \citet{2022PhRvD.105l3501K} \citep[also see][]{2023arXiv230202925K} and as derived in Appendix~\ref{app:2pt},
it is convenient to use the multipole moments of the cross-correlation function 
using the associated Legendre polynomials with $m=2$, denoted as ${\cal L}_\ell^2$: 
\begin{align}
\xi_{\rm g+}(r_p,r_\parallel)\equiv \sum_{\ell\ge 2}\xi_{\rm g+}^{(\ell)}(r){\cal L}^2_{\ell}(\mu_{\bf r}),
\end{align}
where $\mu_{\bf r}$ is the cosine angle between ${\bf r}$ and the line-of-sight direction and 
$\xi^{(\ell)}_{g+}$ is the $\ell$-th order multipole moment. 
Note that the multipole index $\ell$ starts from $2$ ($\ell=2,3,\dots$) and  $\mathcal{L}_2^2(x)=3(1-x^2)$, $\mathcal{L}_4^2(x)=15(1-x^2)(7x^2-1)/2$, and so forth.
The multipole moments $\xi^{(\ell)}_{g+}$ can also be expressed in terms of the cross power spectrum using 
%
\begin{align}
\xi^{(\ell)}_{\rm g+}(r)=
i^{\ell}\int\!\frac{k^2\mathrm{d}k}{(2\pi)^2}P^{(\ell)}_{\rm gE}(k)j_\ell(kr),
\label{eq:xig+_ell_def}
\end{align}
where $P^{(\ell)}_{\rm gE}(k)$ is the corresponding multipole moments of the IA cross power spectrum $P_{\rm gE}({\bf k})$. 

Assuming the linear alignment model (Eq.~\ref{eq:LA}) and the linear Kaiser RSD, the cross-power spectrum is given as
\begin{align}
P_{\rm gE}({\bf k},z)=
b_g b_K\frac{(1-\mu_{\bf k}^2)}{2}(1+\beta \mu_{\bf k}^2) 
P^{\rm NL}_{\rm mm}(k,z),
\label{eq:PgE_def}
\end{align}
where $b_K$ is the linear shape bias parameter \citep{2014PhRvD..89h3507S,2021MNRAS.501..833K,2021JCAP...04..041A}, $b_g$ is the linear bias parameter of the density sample, $\beta\equiv f(z)/b_g$, 
$f$ is the logarithmic of linear growth rate, 
and $\mu_{\bf k}$ is the cosine angle between ${\bf k}$ and the LOS direction. In In $\Lambda$CDM cosmology/Universe, for a wide range of redshifts, $f(z) \sim \Omega_m(z)^{0.55}$.
In the above equation, we used the nonlinear matter power spectrum, $P^{\rm NL}_{mm}$, including the effect of nonlinear structure formation, which is the so-called
nonlinear alignment model (NLA) \citep{2007NJPh....9..444B}. 
Also note that we assumed the linear Kaiser RSD factor $(1+\beta\mu^2)$, but we will below consider the projected correlation function to minimize the RSD contribution.
The shape bias parameter $b_K$ is related to the IA amplitude parameter $A_{\rm IA}$ that is often used in the literature as
\begin{align}
b_K=-2A_{\rm IA}C_1\rho_{\rm crit}\frac{\Omega_{\rm m}}{D(z)},
\end{align}
where $D(z)$ is the linear growth factor and we take $C_1\rho_{\rm crit}=0.0134$ following the convention 
\citep{2011A&A...527A..26J}. Throughout this paper we focus on $A_{\rm IA}$ to discuss the IA amplitude of redMaPPer clusters.

Using Eq.~(\ref{eq:PgE_def}), 
the multipole moments of the cross-correlation function can be found, as derived in 
Appendix~\ref{app:2pt}, as
\begin{align}
&\xi^{(2)}_{g+}(r)=\frac{b_g b_K}{6}\left(1+\frac{\beta}{7}\right)\xi^{(2)}_{\rm mm}(r), \nonumber\\
&\xi^{(4)}_{g+}(r)=\frac{b_g b_K}{105}\beta \xi^{(4)}_{\rm mm}(r),
\label{eq:IAxi_multipoles_def}
\end{align}
and zero otherwise. The multipole moments of the matter two-point correlation function is defined similarly to Eq.~(\ref{eq:xig+_ell_def}) using $P_{\rm mm}^{\rm NL}$.
When there is no RSD effect, only the lowest order moment ($\ell=2$) carries all the IA cross-correlation information, which can be realized by the use of the associated Legendre polynomials \citep{2022PhRvD.105l3501K}.

In this paper we consider the projected IA cross-correlation function defined as
\begin{align}
w_{\rm g+}(r_p)=2\int dz W(z) \int^{\Pi_{\rm max}}_0\!\mathrm{d}r_\parallel~ \xi_{\rm g+}(r_\parallel,r_p; z).
\label{eq:wg+_def}
\end{align}
We adopt $\Pi_{\rm max}=100~\mpc$ as our fiducial choice.

To estimate the linear bias parameter of the density sample, $b_g$, we model the galaxy clustering signal using
\begin{equation}
\label{eq:wgg_model}
    w_{\rm gg}(r_p) = 2\int dz W(z) f_{\rm corr}(r_p, z)\int^{\Pi_{\rm max}}_{0} dr_\parallel b_g^2\xi^{\rm NL}_{\rm mm}(\sqrt{r_p^2+r^2_\parallel}, z), 
\end{equation}
where $f_{\rm corr}(r_p, z)$ is Kaiser correction factor given by \citep{2013MNRAS.430..725V}, 
\begin{equation}
f_{\rm corr}(r_p, z) = \frac{\int_0^{\Pi_{\rm max}}\xi_{\rm gg}^{\rm lin}(r_p, r_\parallel, z)dr_\parallel}{\int_0^{\Pi_{\rm max}}\xi_{\rm gg}^{\rm lin}(\sqrt{r_p^2+r^2_\parallel}, z)dr_\parallel}.
\end{equation}
$\xi_{\rm gg}^{\rm lin} (r_p, r_\parallel, z)$ and $\xi_{\rm gg}^{\rm lin}(r\equiv \sqrt{r_p^2+r_\parallel^2}, z)$ here are the linear two-point galaxy correlation function in redshift space and real space, respectively, where $\xi_{\rm gg}^{\rm lin}(r, z)= b_g^2 \xi_{\rm mm}^{\rm lin}(r, z)$ and the linear galaxy correlation function in redshift space is 
\begin{equation}
\xi_{\rm gg}^{\rm lin}(r_p, r_\parallel, z) = \displaystyle\sum_{l=0}^{2} \xi_{2l}(s, z) \mathcal{P}_{2l} (\mu).
\end{equation}
$s=\sqrt{r_p^2+r_\parallel^2}$ is the real space separation, $\mu = r_\parallel/s$, and $\mathcal{P}_{2l} (x)$ is the $l$th Legendre polynomial. $\xi_0$, $\xi_2$, and $\xi_4$ are given by 
\begin{equation}
\xi_0(r, z) = (1+\frac{2}{3}\beta + \frac{1}{5}\beta^2) \xi_{\rm gg}^{\rm lin}(r, z),
\end{equation}

\begin{equation}
\xi_2(r, z) = (\frac{4}{3}\beta + \frac{4}{7}\beta^2)\left[ \xi_{\rm gg}^{\rm lin}(r, z) - 3J_3(r, z) \right],
\end{equation}
\begin{equation}
\xi_4(r, z) = \frac{8}{35}\beta^2 \left[ \xi_{\rm gg}^{\rm lin}(r, z) + \frac{15}{2} J_3(r,z) - \frac{35}{2}J_5(r, z) \right],
\end{equation}
where
\begin{equation}
    J_n(r, z) = \frac{1}{r^n}\int^{r}_0 \xi_{\rm gg}^{\rm lin}(y, z) y^{n-1}dy.
\end{equation}

To compute the model predictions of the  projected IA cross correlation (Eq.~\ref{eq:wg+_def}), 
We assume the $\Lambda$CDM cosmology with $\Omega_{\rm DM}=0.236, \Omega_b = 0.046, \Omega_{\Lambda} = 0.718, n_s = 0.9646, \sigma_8 = 0.817, h = 0.7$ (WMAP9 cosmology, \citealt{2013ApJS..208...19H}). For the nonlinear matter power spectrum, we employ {\tt Halofit}\footnote{https://pyhalofit.readthedocs.io/} for the $\Lambda$CDM model \citep{2012ApJ...761..152T}. 
We vary the linear bias parameters $b_g$ and $b_K$ (equivalently $A_{\rm IA}$) and estimate the best-fit values by comparing the model predictions with the measurements for the $\Lambda$CDM model.

\section{Data}
\label{sec: data}

\begin{table*}
\centering
\caption{Summary of the sample properties. For the samples in mock observe catalogue, we select clusters based on observed richness $\lambda_{\rm obs}$ with $\gamma_{\rm obs}$; while for mock true, we use true richness $\lambda_{\rm true}$ with $\gamma_{\rm true}$. $N_g$ and $N_{\rm clus}$ are number of galaxies and clusters in the samples separately. $\langle{\lambda}\rangle$ is the mean richness parameter of the sample. $\epsilon_{\rm RMS}^2=\langle \epsilon_i^2 \rangle$ is the RMS ellipticity. $b_g$ and $b_{\rm clus}$ are bias of the samples. The error bars of $b_{clus}$ in the mock indicate the $1\sigma$ scatter among the $19$ mock realizations. $A_{\rm IA}$ is the IA strength parameter obtained from fitting with NLA model. The error bars of $A_{\rm IA}$ indicate the $1\sigma$ scatter among the $19$ mock realizations.}
\begin{tabular}{l|c|c|c|c|c} \hline \hline
\textbf{Observation Dataset} & $N_g$  & $\langle{\lambda}\rangle$ & $\epsilon_{\rm RMS}$  & $b_{\rm g}$  & $A_{\rm IA}$ \\
\hline
LOWZ galaxy & 239,904 & - & - & $1.73\pm0.05$ & - \\
cluster w/ BCG shape ($\lambda\ge20$) & 4,325 & 33.1  & 0.20 & $3.93\pm0.29$ & $11.5\pm3.9$ \\
cluster ($\lambda\ge20$) & 6,345  & 33.0 & 0.21 & $4.69\pm0.25$ & $17.4\pm3.7$ \\
cluster ($20\le\lambda<30$) & 3,593 & 24.2 & 0.22 & $4.12\pm0.36$ & $17.8\pm5.0$ \\
cluster ($30\le\lambda<40$) & 1,492 & 34.3 & 0.20 & $4.73\pm0.64$ & $16.9\pm6.7$ \\
cluster ($40\le\lambda<55$) & 786 & 46.4 & 0.19 & $5.53\pm1.18$ & $10.9\pm7.9$ \\
cluster ($55\le\lambda<200$) & 474 & 73.0 & 0.18 & $6.46\pm1.46$ & $24.5\pm12.1$ \\

\hline
\hline
\textbf{Mock observe} & $\langle N_{\rm clus}\rangle$ &  $\langle \lambda_{\rm obs}\rangle$ & RMS $\epsilon_{\rm obs}$ & $b_{\rm clus}$ & $A_{\rm IA}$ \\
\hline
halos ($M_h>10^{12}\msun$) & - & - & - & $1.17\pm0.02$ & - \\
cluster ($20\le\lambda<200$) & 11,447 & 32.3 &  0.23 & $3.79_{-0.06}^{+0.08}$ & $14.0_{-0.9}^{+0.6}$ \\
cluster ($20\le\lambda<30$) & 7,002 & 24.0 & 0.24 & $3.35_{-0.13}^{+0.07}$ & $12.7_{-0.8}^{+0.6}$ \\
cluster ($30\le\lambda<40$) & 2,328 & 34.2 & 0.22 & $4.04_{-0.32}^{+0.13}$ & $14.7_{-1.8}^{+0.8}$ \\
cluster ($40\le\lambda<55$) & 1,278 & 46.1 & 0.20 & $4.45_{-0.43}^{+0.36}$ & $16.7_{-2.3}^{+2.4}$ \\
cluster ($55\le\lambda<200$) & 839 & 75.3 & 0.19 & $5.56_{-0.62}^{+0.45}$ & $19.6_{-1.4}^{+1.7}$ \\

\hline
\hline
\textbf{Mock true} & $\langle N_{\rm clus}\rangle$ &  $\langle \lambda_{\rm true}\rangle$ & RMS $\epsilon_{\rm true}$ & $b_{\rm clus}$ & $A_{\rm IA}$  \\
\hline
cluster ($20\le\lambda<200$) & 12,848 & 33.7 & 0.32 & $3.28_{-0.09}^{+0.10}$ & $37.3_{-0.3}^{+0.5}$ \\
cluster ($20\le\lambda<30$) & 7,329 & 23.6 & 0.34 & $2.84_{-0.14}^{+0.15}$ & $34.2_{-0.6}^{+1.6}$ \\
cluster ($30\le\lambda<40$) & 2,673 & 33.8 & 0.31 & $3.42_{-0.31}^{+0.19}$ & $37.6_{-2.7}^{+1.3}$ \\
cluster ($40\le\lambda<55$) & 1,590 & 45.8 & 0.30 & $3.68_{-0.22}^{+0.37}$ & $41.6_{-1.6}^{+2.2}$ \\
cluster ($55\le\lambda<200$) & 1,255 & 77.9 & 0.29 & $4.86_{-0.28}^{+0.27}$ & $47.5_{-4.1}^{+1.8}$ \\
\hline

\end{tabular}
\label{tab_cluster}
\end{table*}

\subsection{BOSS DR12 LOWZ Galaxies}
We use SDSS-III BOSS DR12 LOWZ galaxies with spectroscopic redshifts in the range of $0.1\le z\le0.33$ as a biased tracer of the matter field. This is due to their significant overlap with $\redmapper$ clusters. The LOWZ sample consists of luminous red galaxies at $z<0.4$, selected from the SDSS DR8 imaging data and observed spectroscopically in the BOSS survey. The sample is roughly volume limited in the redshift range $0.16<z<0.36$ and has a mean number density of $\sim 3\times 10^{-4}h^3{\rm Mpc^{-3}}$.
We utilize the large-scale structure catalogues\footnote{https://data.sdss.org/sas/dr12/boss/lss/} for BOSS \citep{2012MNRAS.427.3435A, 2016ApJS..224....1R}. 
Table~\ref{tab_cluster} provides an overview of the properties of the density sample. The final density sample contains $239,904$ galaxies. We apply a weighting scheme to sample, using $w=w_{\rm FKP}\times w_{\rm tot}$, where $w_{\rm tot}=w_{\rm sys}\times (w_{\rm cp}+w_{\rm noz}-1)$ for density data and $w=w_{\rm FKP}$ for density random.

\subsection{$\redmapper$ Cluster}
We use galaxy clusters identified with $\redmapper$ algorithm \citep{2014ApJ...783...80R,2014ApJ...785..104R} on SDSS DR8 photometry data \citep{2011ApJS..193...29A}, over an area of about $10,000\ {\rm deg^2}$. The $\redmapper$ algorithm finds optical clusters via identifying overdensity of red sequence galaxies. We use the publicly available version, v6.3. 
For each cluster, the algorithm provides potential brightest central galaxy (BCG) candidates, cluster richness $\lambda$ which is the sum-up of $p_{\rm mem}$ over all candidates members, photometric redshift $z_{\lambda}$, and spectroscopic redshift $z_{\rm spec}$ if available. $p_{\rm mem}$ gives the membership probability of each galaxy belonging to a cluster in the redMaPPer catalog. We choose the galaxies with the highest $p_{\rm cen}$ as BCGs. 
In this paper we use galaxy clusters that have available $z_{\rm spec}$, and select clusters with $20\leq\lambda\leq200$ and $0.1\leq z_{\rm spec}\leq 0.33$. We further divide the sample into sub-samples with $20\le\lambda<30$, $30\le\lambda<40$, $40\le\lambda<55$, $55\le\lambda<200$, in order to study the richness dependence of $A_{\rm IA}$.  The statistical properties of the $\redmapper$ clusters are summarized in Table~\ref{tab_cluster}. 

We use the public random catalog of $\redmapper$ cluster, which includes cluster positions, redshift, richness $\lambda$ and weight. The weighted $z$ and $\lambda$ distributions are the same as in the data. We apply the same $z$ and $\lambda$ cuts in the random catalog for each cluster sample.

\subsubsection{Cluster shape characterization -- BCG versus member galaxy distribution}

We quantify the shape of each $\redmapper$ cluster by two ways: the shape of BCGs, and 
the distribution of the member galaxies relative to BCGs. The BCG shape can be obtained by cross matching with SDSS DR8 shear catalog \citep{2012MNRAS.425.2610R}. 
$4,325$ clusters have BCG shape measurement, out of $6,345$ selected clusters with $20\leq\lambda\leq200$.

Alternatively, we follow the method in \citep{2017MNRAS.468.4502V} to quantify
the cluster shape using member galaxy positions with respective to the BCG.
Using all cluster members with $p_{\rm mem}>0.2$, the second moments of the projected shape are given as
\begin{equation}
\label{eq_rm_mem}
I_{ij}=\frac{\sum_k (\theta_{i, k}-\theta_i^{\rm BCG})(\theta_{j, k}-\theta_j^{\rm BCG})p_{{\rm mem}, k}}{\sum_k p_{{\rm mem}, k}},  
\end{equation}
where $i,j \in {1, 2}$.

The ellipcitity components are then defined as
\begin{equation}
\epsilon_1 = \frac{I_{11}-I_{22}}{I_{11}+I_{22}}, \, \, 
\epsilon_2 = \frac{2I_{12}}{I_{11}+I_{22}}.
\end{equation}
The ``shear'' of cluster shape is estimated as
$\gamma_{1,2}=\epsilon_{1,2}/(2{\cal R})$,
where $\mathcal{R} \equiv 1-\langle \epsilon_i^2\rangle$ is the shear responsivity \citep{2002AJ....123..583B}.

\subsection{Correlation Function Estimator} 
\label{sec: corr_func}

For the BOSS LOWZ sample and the specp-$z$ matched redMaPPer cluster, we measure the auto-correlation function
of LOWZ galaxies, $\xi_{\rm gg}({\bf r})$,  
and the projected IA cross-correlation function between the LOWZ galaxy and the redMaPPer cluster shapes, 
$\xi_{\rm g+}({\bf r})$.

We use a generalized Landy-Szalay estimator \citep{1993ApJ...412...64L} for estimating the correlation functions: 
\begin{align}
    \hat{\xi}_{g+} &= \frac{S_+ D - S_+ R_D}{R_S R_D}, \\
    \hat{\xi}_{gg}&=\frac{DD-2DR+RR}{RR},
\end{align}
where $S_+$ is the shape field for the cluster sample, $D$ is the density field for the LOWZ galaxy sample, 
and $R_S$ and $R_D$ are random points corresponding to shape sample and density sample, respectively. 
$S_+$ is the $+$-component of cluster shear with respect to the vector ${\br}\equiv{\bf x}-{\bf x}'$ connecting the cluster position 
and the LOWZ galaxy or the density random point (see Eq.~\ref{eq:shear+_def}).

For the IA cross-correlation, we consider the projected correlation function:
\begin{equation}
    \hat{w}_{\rm g+} (r_p) = \int^{\Pi_{\rm max}}_{-\Pi_{\rm max}}\!\mathrm{d}\Pi~ \hat{\xi}_{\rm g+} (r_\parallel,r_p).
\end{equation}
We compare the measured $w_{g+}$ with the theory prediction (Eq.~\ref{eq:wg+_def}).

\subsection{redMaPPer Cluster Mock}
To study the impact of projection effects on IA of galaxy clusters, we use the cluster mock catalog constructed in \citet{2019MNRAS.490.4945S} (see also \citealt{2020MNRAS.496.4468S} and \citealt{2022arXiv220503233S}). Here we briefly summarize the mock construction procedures, and refer the readers to \citet{2019MNRAS.490.4945S} for more detailed information.

To construct the cluster mock, $N$-body simulations from \citet{2019ApJ...884...29N} are used, which were performed with $2048^3$ particles in a comoving cubic box with side length of $1\gpc$. The simulations adopt
the $Planck$ Cosmology \citep{2016A&A...594A..13P}. 
The particle mass is $1.02\times 10^{10}\msun$. Halos are identified using Rockstar halo finder \citep{2013ApJ...762..109B}, and $M_{200m}$ is adopted for halo mass, which is the total mass within $R_{200m}$. $R_{200m}$ is the radius within which the mean density is $200$ times the mean mass density $\bar{\rho}_m$. For our purpose, we use the simulation snapshot and halo catalogs at $z=0.25$, which is the mean redshift of the $\redmapper$ clusters. We have $19$ realizations of $N$-body simulation and cluster mock.

Mock galaxies are populated into halos with mass $M_{200m}>10^{12}\msun$ using halo occupation distribution (HOD) prescription \citep{2005ApJ...633..791Z}. 
The HOD parameters are chosen to match with the abundance and lensing measurements of the $\redmapper$ clusters.
Instead of distributing the satellite galaxies using Navarro-Frenk-White profile \citep{1997ApJ...490..493N}, the satellites are populated using the positions of randomly selected member particles in each halo.
As a result, the satellites distribution within the halo traces the non-spherical halo shape, which is also used as one of the validation tests in Appendix~\ref{append: mock_varying_shape}.

The photometric redshift uncertainty, which is the main source of the projection effects, is modeled by assuming a specific projection length, $d_{\rm proj}$. In this work, we use the mock with $d_{\rm proj}=60\mpc$.
The cluster finder which mimics the $\redmapper$ algorithm \citep{2014ApJ...783...80R, 2014ApJ...785..104R} is then run on the red-sequence mock galaxies, producing the mock cluster catalog that includes the true richness $\lambda_{\rm true}$, the observed richness $\lambda_{\rm obs}$, and the membership probability $p_{\rm mem}$. The galaxy in the most massive halo in each identified cluster is considered as the central galaxy of the cluster. The optical radial cut that scales with the richness, $R_c(\lambda) = R_0 (100/\lambda)^{\beta}$, is applied the same way as in observation when running the $\redmapper$ algorithm in mock, where $R_0=1.0\mpc$ and $\beta=0.2$.

Similar as in observation, we divide the mock cluster sample into subsamples with various richness bins, using both $\lambda_{\rm obs}$ and $\lambda_{\rm true}$. We use halos with $M_{200m}>10^{12}\msun$ as density tracers, $\delta_h$, of the matter field, where $\delta_h \equiv \frac{n_h(\bf{x})-\overline{n}_h}{\overline{n}_h}$. The properties of the selected cluster samples are shown in Table~\ref{tab_cluster}. The cluster bias increases with the richness, which is consistent with the fact that halo/cluster mass increases with richness. \citet{2020MNRAS.496.4468S} presented the halo mass distribution of the mock clusters in different richness bins (divided by both $\lambda_{\rm obs}$ and $\lambda_{\rm true}$), showing that mass distributions for the \textbf{``mock observe"} sample is more extended than the \textbf{``mock true"} sample because of the projection effects, also the peak mass shifts towards higher masses from finite aperture effects in higher richness bins. This explains the higher cluster bias for \textbf{``mock observe"} sample shown in Table~\ref{tab_cluster}.

\subsubsection{Cluster shape characterization}
For each galaxy cluster in the mock, we calculate the observed cluster shape $\gamma_{\rm obs}$ using the $\redmapper$ member galaxies with $p_{\rm mem}>0.2$, using Eq.~(\ref{eq_rm_mem}). 
Unlike observation, mock cluster catalogs provide the true positions of the satellite galaxies as well as the dark matter particles.
So, we can calculate the intrinsic cluster shear $\gamma_{\rm true}$ using satellite galaxy positions and $\gamma_{\rm DM}$ using DM particles distributions (see Appendix~\ref{append: mock_varying_shape} for details of the calculation). The IA signal measured from $\gamma_{\rm true}$ agrees with that from $\gamma_{\rm DM}$ very well (see Appendix~\ref{append: mock_varying_shape}). So in the following, we take mock clusters selected using $\lambda_{\rm true}$ and shape calculated using 
 $\gamma_{\rm true}$ as the \textbf{``mock true"} sample, while mock cluster selected using $\lambda_{\rm obs}$ and $\gamma_{\rm obs}$ as the \textbf{``mock observe"} sample.

We use TreeCorr \citep{2004MNRAS.352..338J} to compute the correlation functions. We measured the signal as a function of transverse comoving separation in $25$ logarithmic bins between $0.1$ and $200\mpc$. 
We take $\Pi_{\rm max}=100\mpc$ and $20$ linear bins for $r_\parallel \in[-100, 100] \mpc$.
To estimate the covariance matrix, we divide the \redmapper Cluster sample into $50$ jackknife regions of approximately equal area on the sky, and compute the cross-correlation function by excluding one region each time \citep{2009MNRAS.396...19N}. For the mock cluster sample, we divide the simulation box into $64$ sub-boxes of equal volume for jackknife covariance matrix estimation. 

We restricted the analysis to mildly non-linear scales of $r_p>6\mpc$. The size of the jackknife patch is $14$~deg., which roughly corresponds to $70\mpc$ at $z=0.1$. So we take $70\mpc$ as the maximum scale in the fitting.

\section{Results}
\label{sec: results}

\subsection{IA of $\redmapper$ Clusters in SDSS}
\begin{figure*}
    \centering
    \includegraphics[width=.8\linewidth]{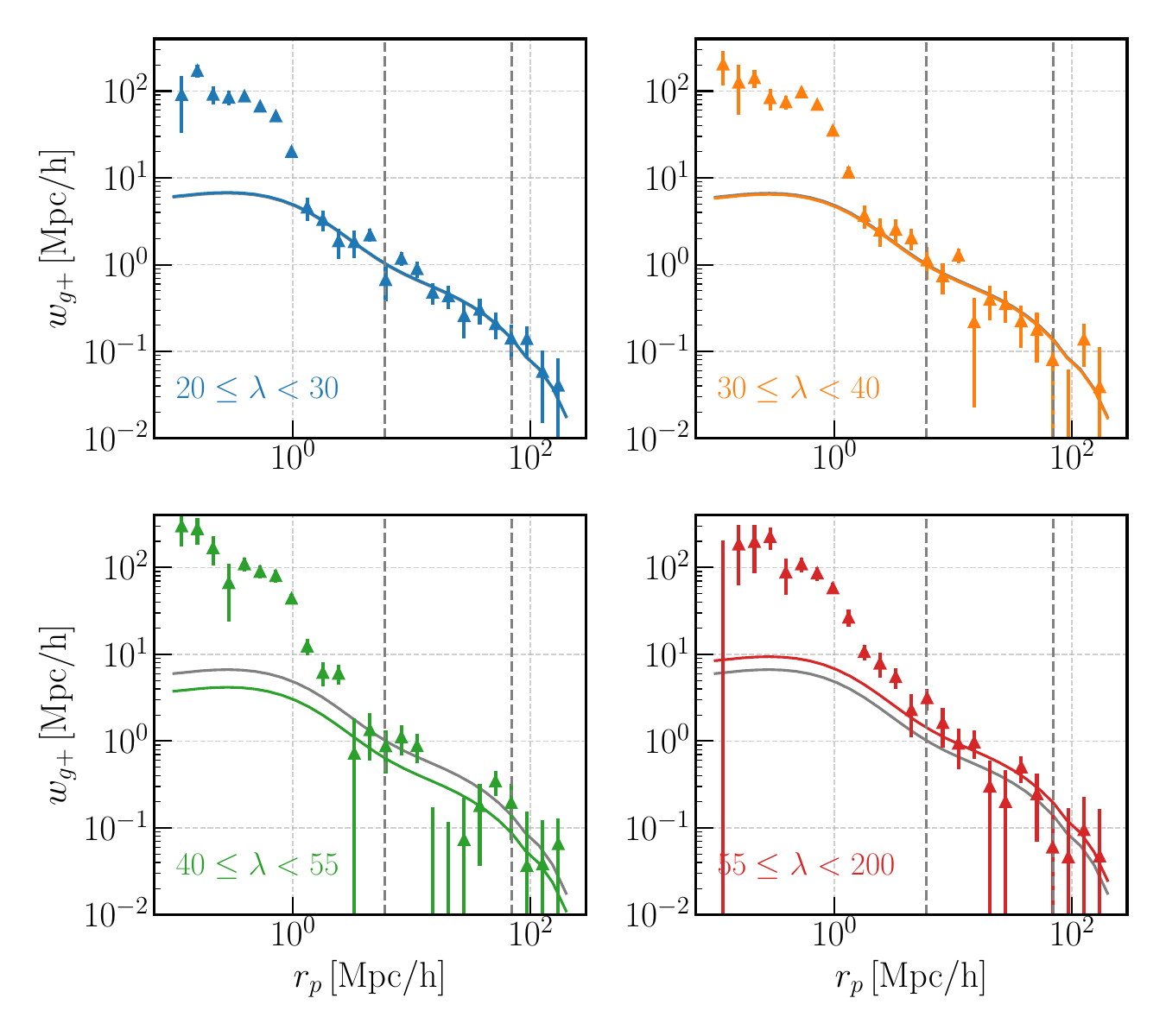}
    \caption{The LOWZ galaxy-cluster shape correlation, $w_{g+}$, of clusters in the redshift range of $0.1\leq z\leq 0.33$. Different panels show the result for various richness bins. The dots are measurement from data, and the colored solid lines are the fitting using NLA in the range of $6\mpc<r_p<70\mpc$. The gray solid line shows the fitting result for the cluster sample with $20\le\lambda<200$, just to guide the eyes.}
    \label{fig:wgp_cluster}
\end{figure*}

The measured cross correlation functions of the galaxy density field and the cluster shape field are shown in Figure~\ref{fig:wgp_cluster}. Here we used the cluster shapes measured using positions of the member galaxies relative to the BCG in each cluster. We obtain a clear detection of IA signal in all richness bins, meaning that cluster shapes have correlations with the surrounding large-scale structures.

\begin{figure}
    \centering
    \includegraphics[width=.95\linewidth]{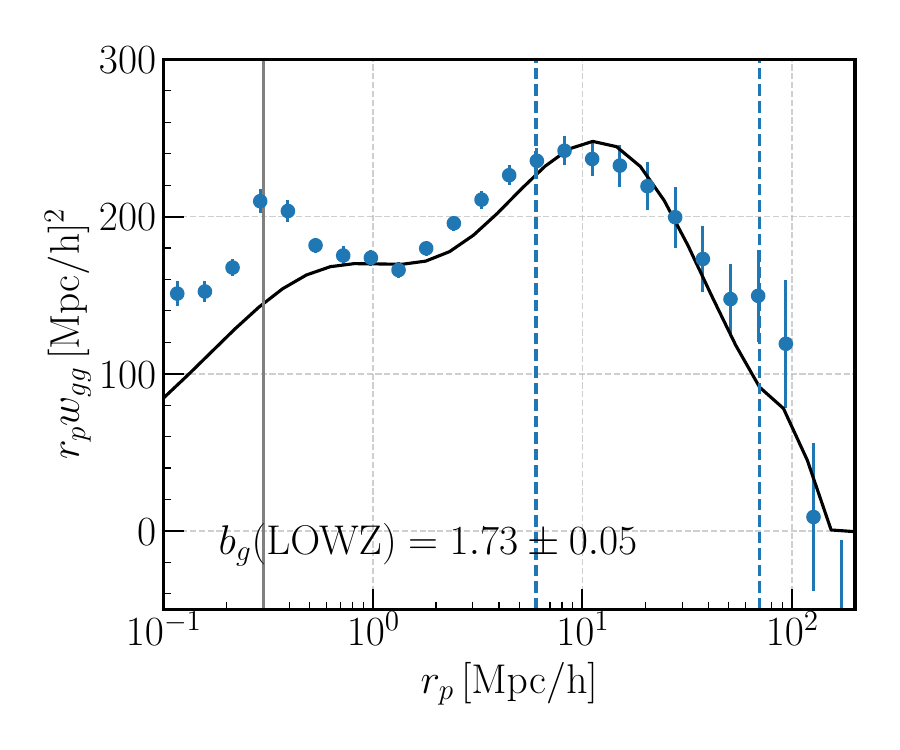}
    \caption{The galaxy-galaxy correlation, $w_{gg}$, of the LOWZ sample in the redshift range of $0.1\leq z\leq 0.33$. The blue dots are measurement from data, and the black solid line is the fitting using linear model with non-linear matter power spectrum in the range of $6\mpc<r_p<70\mpc$. The gray solid line shows the SDSS fibre collision scale at $z=0.33$.}
    \label{fig:wgg_lowz}
\end{figure}

The IA amplitude, $A_{\rm IA}$, is obtained by fitting NLA model to the measurement, as introduced in Section~\ref{sec: method}. 
However, $A_{\rm IA}$ is degenerate with bias parameter $b_g$ of the galaxy density sample. We obtain $b_g=1.73\pm 0.05$ by measuring and fitting the projected clustering signal of LOWZ galaxies to the model (Eq.~\ref{eq:wgg_model}), as shown in Figure~\ref{fig:wgg_lowz}. We have good fits of the model prediction, with reduced $\chi^2$ value of $1.02$. 
Our result for the LOWZ galaxy bias is consistent with the previous measurement, $b_g$=$1.77\pm 0.04$, in 
\citet{2015MNRAS.450.2195S}.
We ascribe the slight difference to the different redshift range, where they used $0.16<z<0.36$ compared to our range, $0.10\le z\le0.33$.

The IA amplitude of each subsample can be found in Table~\ref{tab_cluster}. The NLA model gives a good fit to the measured $w_{g+}$ in the fitting range of $6<r_p<70\mpc$ for each cluster sample. However, at small scales, the model predictions are much lower than the measured signal. The IA amplitude, $A_{\rm IA}$, does not show a clear dependence on cluster richness.
This contradicts with the results found from the shapes of halos in simulations \citep{2021MNRAS.501..833K}; they found that $A_{\rm IA}$ increases with halo mass. We found this is mainly caused by the projection effects, as we will discuss in Section~\ref{sec: lambda_dependence} in detail.

\subsubsection{Tests for systematics}

\begin{figure}
    \centering
    \includegraphics[width=1.\linewidth]{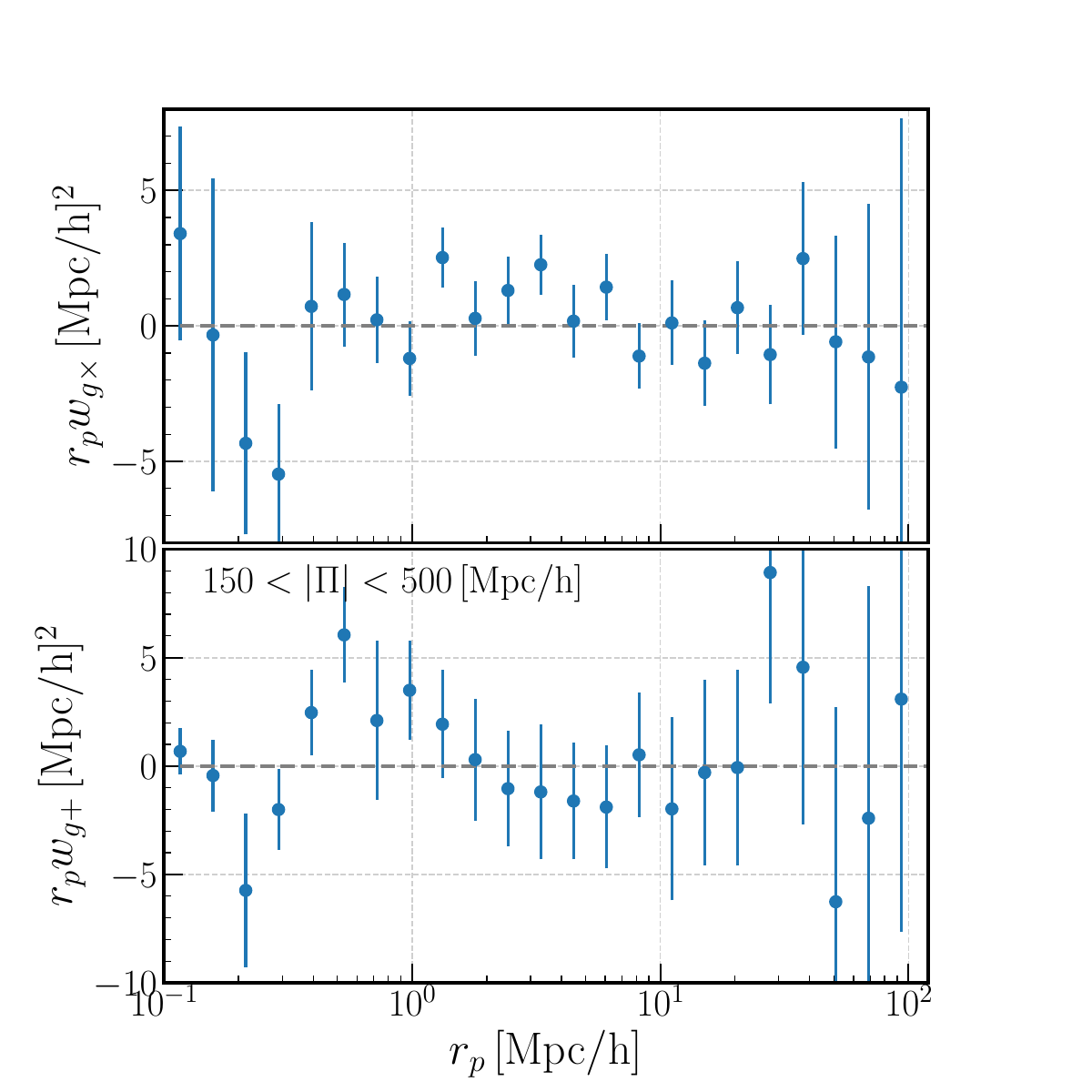}
    \caption{Tests of systematic effects in  the IA measurement of galaxy clusters with $20\le\lambda<200$. {\it Upper panel}: cross correlation between LOWZ galaxies and \redmapper cluster shape $\gamma_{\times}$, the signal is consistent with $0$. {\it Lower}: measurement of $w_{g+}$ within $150<|\Pi|<500\mpc$, the signal is also consistent with null signal.}
    \label{fig:sys_test}
\end{figure}

In Figure~\ref{fig:sys_test} we study potential systematic effects in our IA measurements. The upper panel shows the measured correlation function between the cross-component of the cluster shape, $\gamma_{\times}$, and the galaxy density field, $w_{g\times}$, for the sample with $20\leq \lambda<200$.
This cross correlation should be vanishing due to parity symmetry if the measurements is not affected by an unknown systematic effect.
We also show the IA cross-correlation function, $w_{g+}$, measured by integrating the original 3D IA correlation function only over the large line-of-sight separation, $150<|\Pi|<500\mpc$. This cross-correlation is expected to have a very small signal, if the redshift of clusters is accurate or if there is no significant contamination of fake clusters
due to the projection effect.
The measured $w_{g+}$ for the large $|\Pi|$ separation shows a very small signal. Hence we conclude that our measurements are not affected by the $\times$-component or the fake clusters.

There are other potential systematic effects that affect our IA measurements. These include photometric redshift errors, errors in cluster shape estimation arising due to a limited number of member galaxies, miscentering effect, contamination of merging clusters, and incompleteness of cluster sample or selection function.
\citet{2017MNRAS.468.4502V} presented the tests of above systematic effects for the $\redmapper$ cluster sample, and showed that the most significant systematic effect arises from photo-$z$ errors for the cluster sample.
Since we use only the clusters that have spectroscopic redshifts, we conclude that our IA measurements are not affected by the photo-$z$ errors. 

However, we below show that the projection effect due to large-scale structure surrounding the \redmapper clusters causes a systematic contamination to the IA measurements.

\subsection{IA of Clusters in Mock - Impact of Projection Effect}
\label{sec: mock}
\begin{figure}
    \centering
    \includegraphics[width=.95\linewidth]{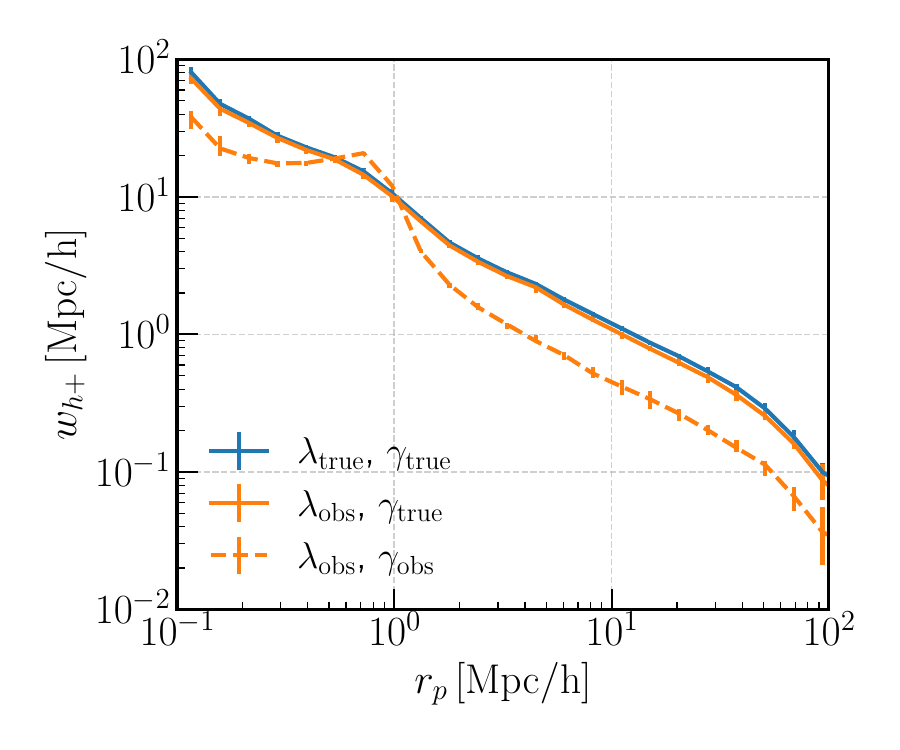}
    \caption{The IA correlation functions measured for galaxy clusters in the mock simulations, selected using $20\le\lambda_{\rm obs}<200$ (blue) and $20\le\lambda_{\rm true}<200$ (orange), respectively.
    The solid lines show the measurements using $\gamma_{\rm true}$, i.e., satellites distributions within dark matter halos, while the dashed line shows the measurement using $\gamma_{\rm obs}$, i.e., member galaxies identified by $\redmapper$ algorithm. The lines here show the median values among $19$ realizations, and the error bars are the $1\sigma$ dispersion.}
    \label{fig:wgp_mock}
\end{figure}

In Figure~\ref{fig:wgp_mock} we study the impact of the projection effect on the IA correlation functions using the mock catalog 
of \redmapper clusters. To do this, we compare the IA correlation functions for clusters using the true or ``observed'' richness ($\lambda_{\rm true}$ or $\lambda_{\rm obs}$) and/or the true or ``observed'' shape estimates ($\gamma_{\rm true}$ or $\gamma_{\rm obs}$), where the observed quantities are affected by the projection effect.
The figure shows that the IA correlation function using the observed quantities ($\lambda_{\rm obs}$ and 
$\gamma_{\rm obs}$) displays about factor of $2$ smaller amplitudes than that for non-contaminated clusters ($\lambda_{\rm true}$ and $\gamma_{\rm true}$).
The solid orange curve shows the result when using the clusters for $\lambda_{\rm obs}$ and $\gamma_{\rm true}$, which show almost similar amplitudes to that for the non-contaminated clusters ($\lambda_{\rm true}$
and $\gamma_{\rm true}$). The comparison tells that the smaller amplitude for the case of ($\lambda_{\rm obs}, \gamma_{\rm obs}$) is caused mainly by the projection effect on the shape measurement ($\gamma_{\rm obs}$ against $\gamma_{\rm true}$).
The $A_{\rm IA}$ values estimated from $w_{h+}$ for the different samples are given in Table~\ref{tab_cluster}. Figure~\ref{fig:wgp_mock} only shows the result for the cluster sample with $20\le\lambda<200$, the measurement and fitting results for other richness bins are shown in Appendix~\ref{append: mock_lambda_bins}.

When comparing the solid and dashed lines in Figure~\ref{fig:wgp_mock}, we notice the existence of a bump in $w_{h+}$ around $r_p \sim 1\mpc$ for the case with projection effects. Here $1~h^{-1}{\rm Mpc}$ roughly corresponds to the aperture size used in the \redmapper\, cluster finder  \citep{2014ApJ...785..104R}.
We will show later that this specific imprint of projection effects is likely caused by the non-member interlopers, which are however identified as cluster members by the \redmapper\, method, and the real member galaxies that are missed by the cluster finder. 

\subsubsection{$f_{\rm true}$ and $f_{\rm miss}$}

\begin{figure*}
    \centering
    \includegraphics[width=.75\linewidth]{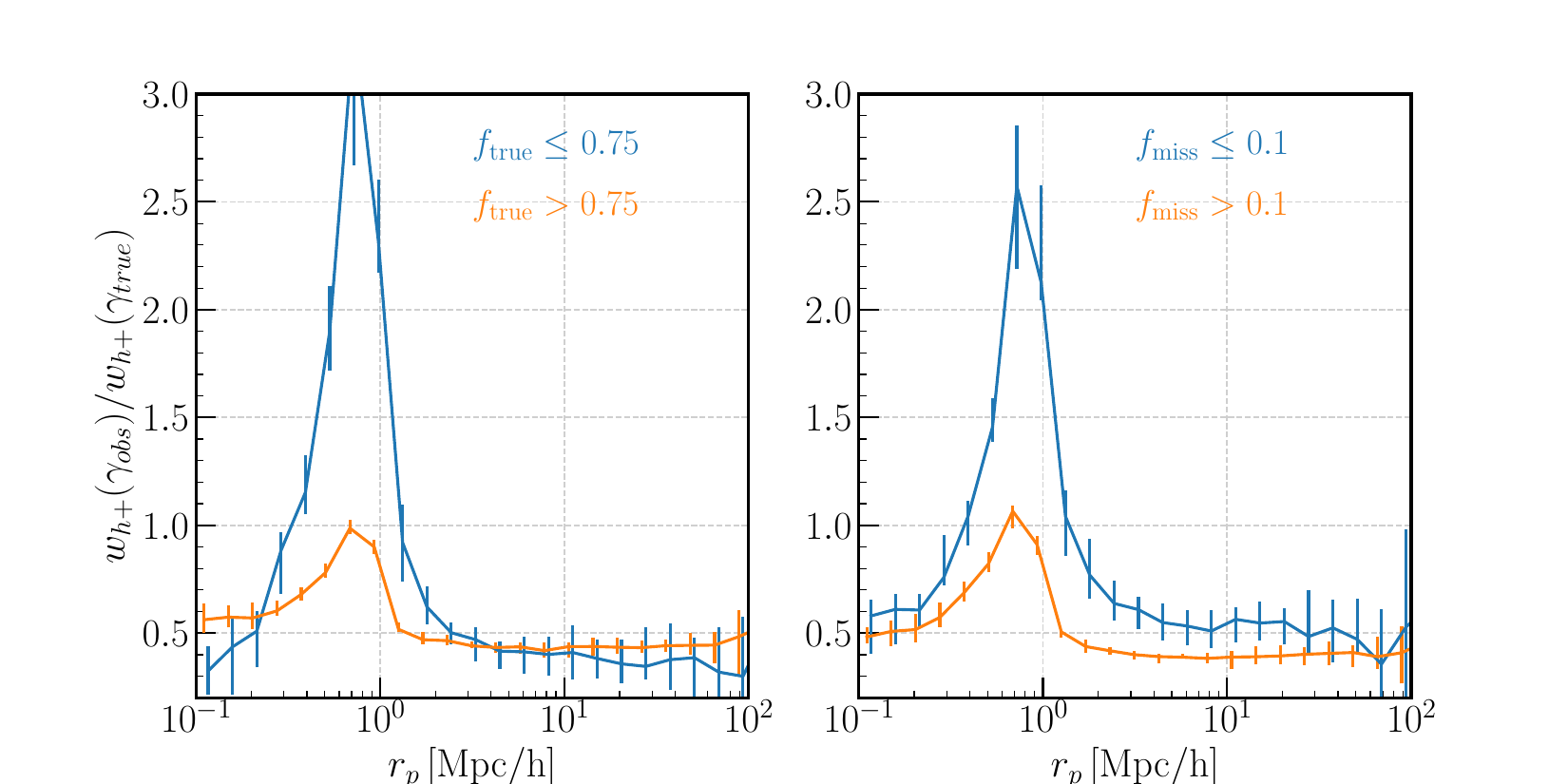}
    \caption{The IA correlation functions measured from the mock cluster catalogs. 
    Shown is the ratio of the IA correlation function using the observed shape ($\gamma_{\rm obs}$) to that of the true shape ($\gamma_{\rm true}$), for a subsample of the mock clusters with $20\le \lambda_{\rm obs}<200$.
    {\it Left panel}: the ratio $w_{h+}(\gamma_{\rm obs})/w_{h+}(\gamma_{\rm true})$ for subsamples with $f_{\rm true}\le 0.75$ and $f_{\rm true}>0.75$, respectively, where $f_{\rm true}$ is the fraction of true members among the cluster members identified by the $\redmapper$ finder in each cluster.  {\it Right}: the ratio for subsamples divided by $f_{\rm miss}\le 0.1$ and $f_{\rm miss}>0.1$, where $f_{\rm miss}$ is the fraction of true members missed by the finder in each cluster.}
    \label{fig:wgp_ratio_ft_fm}
\end{figure*}

As we have found, the projection effect 
impacts the shape estimation of clusters. 
There are two effects: one is caused by including interlopers (non-member galaxies) in the cluster members, and the other
is caused by missing real member galaxies, when estimating the cluster shape.
To study how these two effects cause a contamination to the IA correlation function, we define the following quantities:
\begin{itemize}
    \item $f_{\rm true} = \frac{\sum_{d_i\le R_c} p^{\rm true}_{{\rm mem}, i}}{\lambda_{\rm obs}}$, which is the true member fraction of identified members in each cluster. This quantity is the same as that used in \citet{2020MNRAS.496.4468S},
    \item $f_{\rm miss} = 1. - n_{\rm true, mem}(<R_c)/\lambda_{\rm true}$, which is the fraction of true members missed in the membership identification in each cluster.
\end{itemize}
Here $p_{{\rm mem},i}^{\rm true}$ is the membership probability of the $i$-th true member galaxy identified by the $\redmapper$ 
finder, $R_c$ is the cluster radius used in the $\redmapper$ finder, and $n_{\rm true, mem}$ is the number of true member galaxies among all $\redmapper$ member galaxies. Note
$0<f_{\rm true}\le 1$ by definition, and $f_{\rm true}=1$ means that the $\redmapper$ finder-identified member galaxies are true member galaxies that belong to the cluster, and no interlopers contaminate the true membership (however, all the true members are not necessarily identified). On the other hand, 
a low $f_{\rm true}$ indicates a higher contamination fraction of interlopers. $f_{\rm miss}$ informs how many true member galaxies are not identified as member galaxies by the cluster finder.

In Figure~\ref{fig:wgp_ratio_ft_fm}, we show the ratio of $w_{h+} (\gamma_{\rm obs})$ versus $w_{h+}(\gamma_{\rm true})$ for samples with low $f_{\rm true}$ ($f_{\rm miss}$) and high $f_{\rm true}$ ($f_{\rm miss}$) separately.
If the ratio between $w_{h+}(\gamma_{\rm obs})$ and $w_{h+}(\gamma_{\rm true})$ is close to $1$ for a sub-sample, it means the measured cluster shape/IA are less affected by the projection effects. On contrary, if the ratio deviates from unity more, it means the projection effect is making the measured shape/IA deviates from the underlying true signals. Figure~\ref{fig:wgp_ratio_ft_fm} shows that the impact on large-scale IA signal of projection effects is weaker for clusters with high $f_{\rm true}$ and low $f_{\rm miss}$, compared to the clusters with low $f_{\rm true}$ and high $f_{\rm miss}$. 
The amplitude of the bump at $r_p\sim 1\mpc$ is significantly decreased for samples with higher $f_{\rm true}$ and higher $f_{\rm miss}$. 
As shown in Figure \ref{fig:wgp_mock}, the bump only appears when the projection effect is included in the mock, i.e. for $w_{h+} (\gamma_{\rm obs})$. 

\subsubsection{Coupling between cluster IA and projection effects}
\begin{figure*}
    \centering
    \includegraphics[width=.75\linewidth]{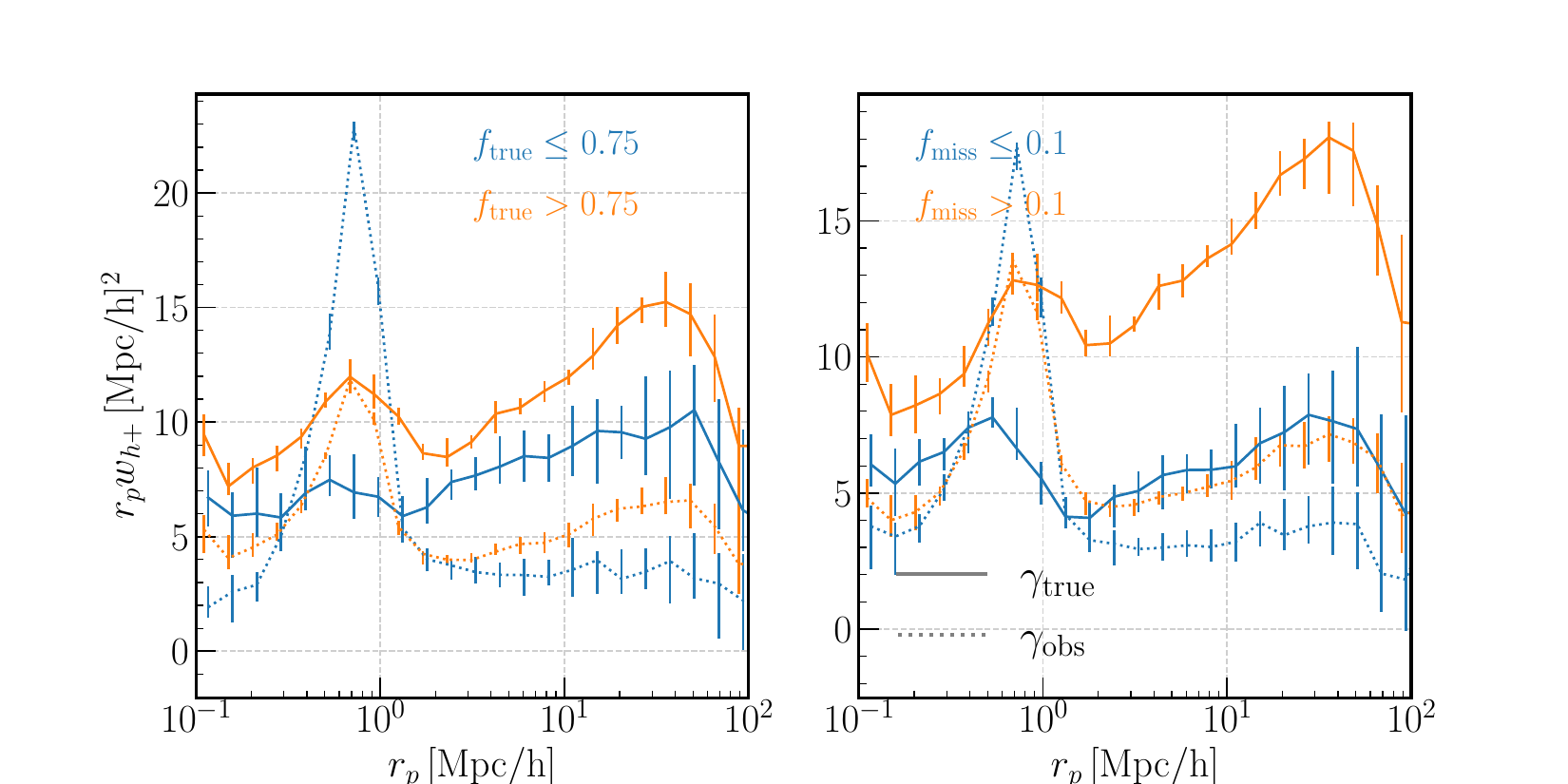}
    \caption{Similar to the previous figure, but we show the IA correlation functions, instead of the ratio. The solid lines are the IA correlation functions measured using $\gamma_{\rm true}$, and the dotted lines are those measured using $\gamma_{\rm obs}$. 
    }    
    \label{fig:wgp_mock_ft_fm}
\end{figure*}

\begin{figure}
    \centering
    \includegraphics[width=1.\linewidth]{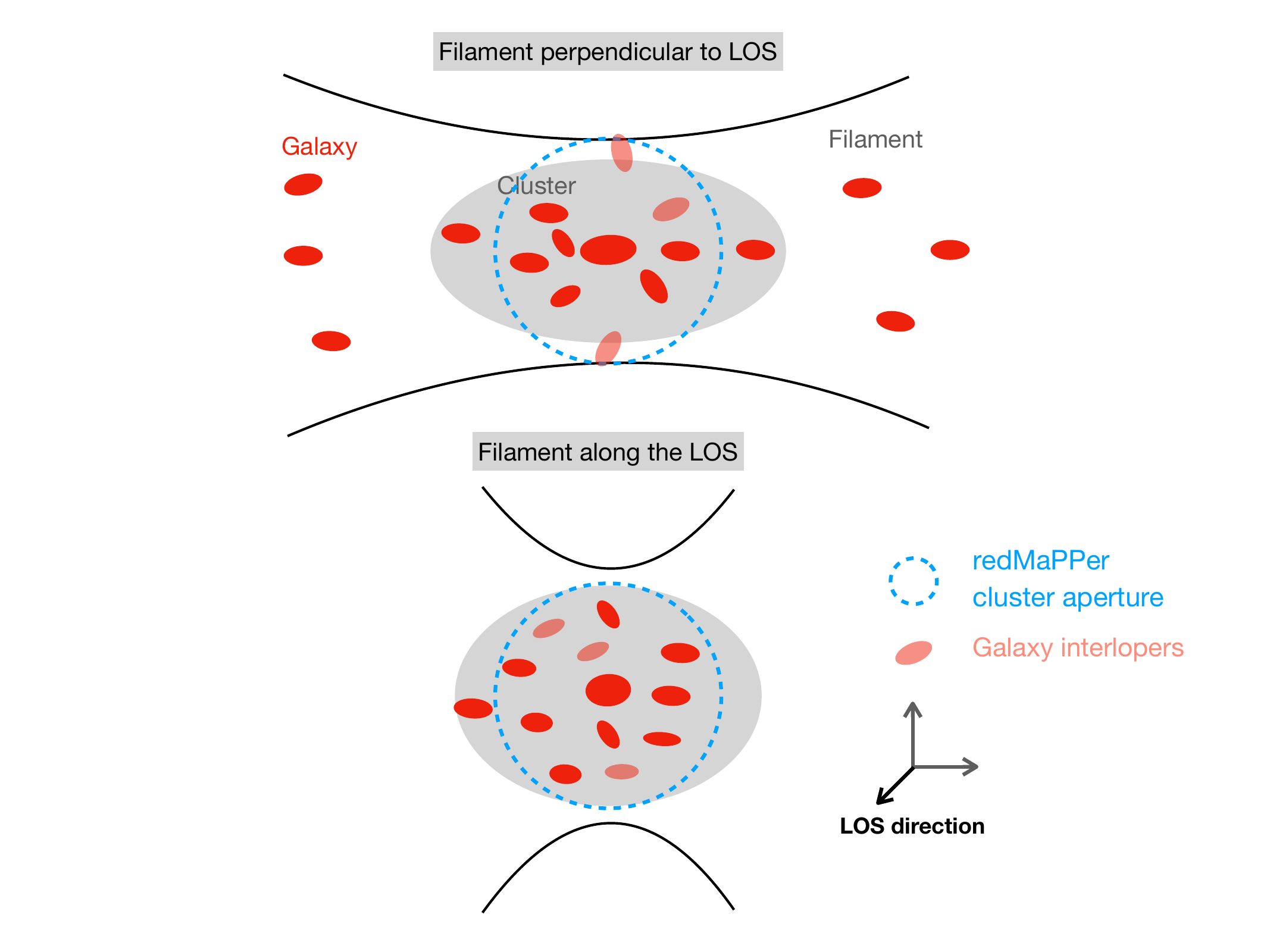}
    \caption{Cartoons illustrating the couplings between galaxy cluster IA and the projection effects, for clusters with their major axis perpendicular to the LOS direction (upper) and along the LOS direction (lower). For clusters with their major axis perpendicular to the LOS, the measured IA signal using projected cluster shape is stronger, the contamination fraction from interlopers is lower (i.e. higher $f_{\rm true}$), and the missed galaxy member fraction is higher (i.e. higher $f_{\rm miss}$); for clusters with their major axis parallel with the LOS direction, the measured IA signal is lower, the contamination from interlopers is more severe (i.e. lower $f_{\rm true}$), and the missed member fraction is lower (i.e. lower $f_{\rm miss}$).}    
    \label{fig:cartoon}
\end{figure}

Cluster IA and projection effects are coupled with each other. In Figure~\ref{fig:wgp_mock_ft_fm}, we compare the IA signal of low $f_{\rm true}$ ($f_{\rm miss}$) and high $f_{\rm true}$ ($f_{\rm miss}$) sub-samples. The large-scale IA amplitude is higher when $f_{\rm miss}$ or $f_{\rm true}$ is higher, for both $\gamma_{\rm obs}$ and $\gamma_{\rm true}$. The coupling between cluster IA and projection effects are illustrated by the cartoons shown in Figure~\ref{fig:cartoon}.

For clusters with their major axis (orientation) perpendicular to the LOS direction, the measured IA is higher, since their projected shapes appear more elliptical and we measured the cross correlation between the projected shapes and the density field. These clusters also tend to have LSS structures, such as filaments, that are perpendicular to the LOS direction. The missed member galaxy fraction $f_{\rm miss}$ is higher, since the projected member galaxies distribution is more dispersed; and the contamination from interlopers is lower, since there are less galaxies outside the cluster along the LOS, thus $f_{\rm true}$ is higher. In contrary, for clusters with their major axis along the LOS direction: the measured IA is lower, and it is more likely to have LSS structures along the LOS; they are less likely to miss galaxy members (lower $f_{\rm miss}$) since they are concentrated in the inner region; the contamination from interlopers along the LOS is higher (lower $f_{\rm true}$). In both cases, the outer region of the cluster is affected more, since the member number density decreases with the distance from the cluster center. This likely explains the existence of the bump at $r_p \sim 1\mpc$, which is also the typical cluster boundary. 
The above picture is supported by Figure~\ref{fig:Pmu} in Appendix~\ref{append: Pmu}, where we show that clusters with lower $f_{\rm true}$ and lower $f_{\rm miss}$ tend to have their major axis parallel with the LOS direction.
In summary, the above picture explains the coupling between cluster IA and $f_{\rm miss}$, $f_{\rm true}$.

\subsection{Dependence on Cluster Richness}
\label{sec: lambda_dependence}

\begin{figure}
    \centering
    \includegraphics[width=.95\linewidth]{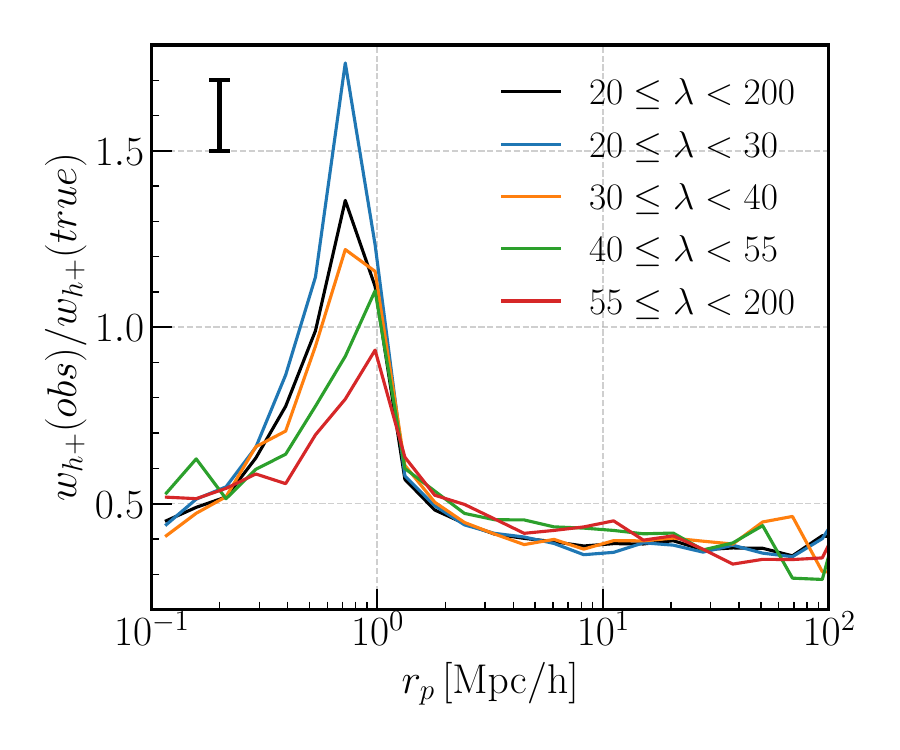}
    \caption{The ratio of the observed IA signal versus the true IA signal of galaxy clusters in various richness bins in the mock. $w_{h+} (obs)$ here is calculated using $\lambda_{\rm obs}$ and $\gamma_{\rm obs}$, and $w_{h+}(true)$ here is calculated using $\lambda_{\rm true}$ and $\gamma_{\rm true}$. The typical $1\sigma$ scatter of the ratio among realizations is shown in the upper left corner of the plot. }
    \label{fig:wgp_mock_ratio}
\end{figure}

The impact of the projection effects on cluster IA is independent of the cluster richness, as shown in Figure~\ref{fig:wgp_mock_ratio}. The ratios of $w_{h+} (obs)$ with projection effects versus $w_{h+}(true)$ without projection effects at scales of $6<r_p<70\mpc$ is roughly constant and doesn't depend on the richness of the clusters. 

\begin{figure}
    \centering
    \includegraphics[width=.95\linewidth]{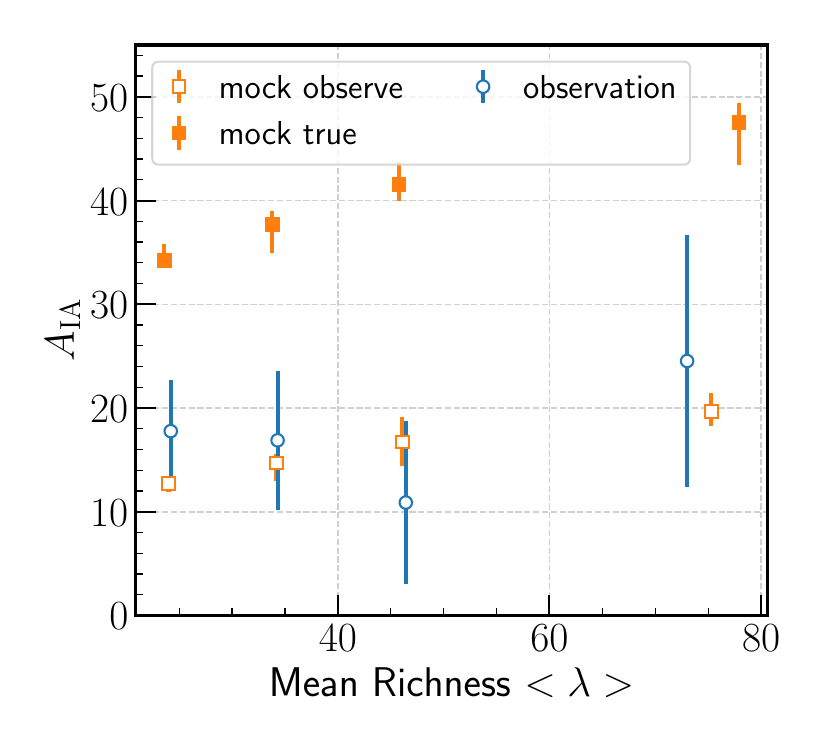}
    \caption{The $A_{\rm IA}$ versus richness $\lambda$ relation for clusters in observation and mock catalog. The blue open circles show the results from fitting the observed $w_{g+}(r_p)$ of $\redmapper$ clusters, the orange open squares are results from mock observe samples using $\lambda_{\rm obs}$ and $\gamma_{\rm obs}$. The orange filled squares are results from mock true samples using $\lambda_{\rm true}$ and $\gamma_{\rm true}$. }
    \label{fig:aia_lambda}
\end{figure}

In Figure~\ref{fig:aia_lambda}, we plot the measured $A_{\rm IA}$ versus cluster mean richness for $\redmapper$ clusters in observation, and clusters in the mock with $w_{h+}(true)$ (filled squares) and $w_{h+}(obs)$ (open squares). The $A_{\rm IA}$ from observation agree with results using $w_{h+}(obs)$ from mock pretty well, indicating that our mock construction and inclusion of the projection effects is quite reasonable. A weak increase of $A_{\rm IA}$ with respect to cluster richness can be seen for clusters free of projection effects. However, such dependence can not be seen once projection effects are included. 
we further derived the $A_{\rm IA}$ - halo mass relation for galaxy clusters and compared it with the prediction from N-body simulation, which is shown in Appendix~\ref{append: aia_mass}.    

\section{Discussion}
\label{sec: discuss}

\subsection{Cluster IA using BCG shape versus member galaxy positions}

\begin{figure}
    \centering
    \includegraphics[width=.95\linewidth]{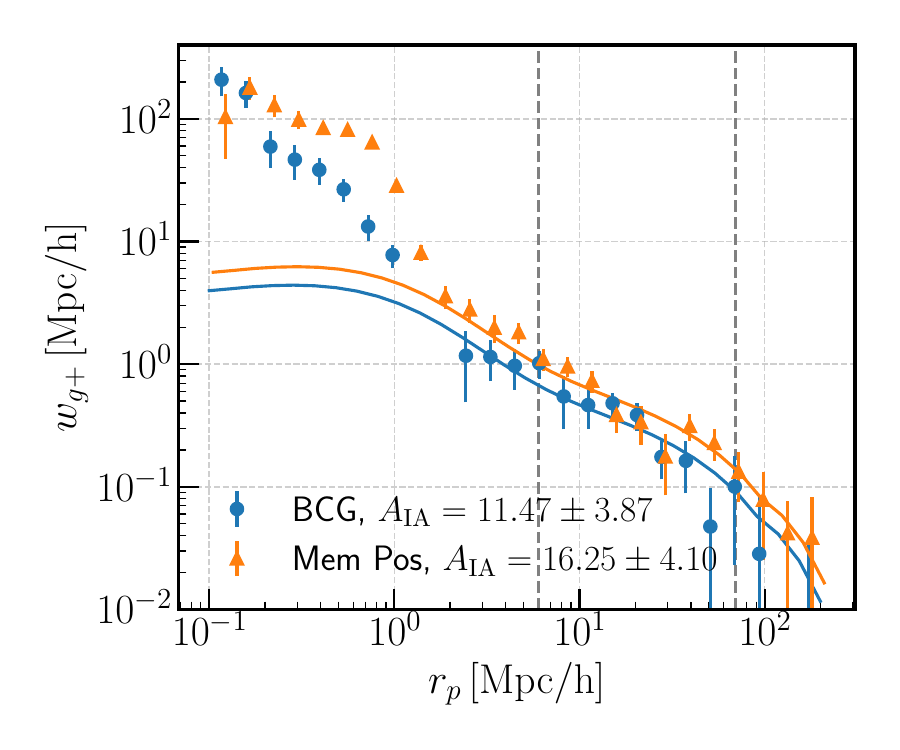}
    \caption{The LOWZ galaxy-BCG shape (blue) and LOWZ galaxy-cluster shape (orange) correlation, $w_{g+}$, of clusters with available BCG shapes in the redshift range of $0.1\leq z\leq 0.33$. The blue and orange dots are measurement using BCG shapes and member galaxy positions, $\gamma_{\rm obs}$, separately. The solid lines are the fitting results using NLA in the range of $6\mpc<r_p<70\mpc$. }
    \label{fig:wgp_bcg}
\end{figure}

The IA of BCGs are shown in Figure~\ref{fig:wgp_bcg}. BCGs show a similar IA amplitude as the clusters that they lie in, indicating the good alignment of BCG orientations with respect to the member galaxies distribution of clusters. If we assume that the member galaxy distributions trace well the dark matter halo shapes, then the results in Figure~\ref{fig:wgp_bcg} could hint a rather good alignment between BCG and dark matter halos.
However, previous studies of \citet{2009ApJ...694..214O} showed that central LRGs are not perfectly aligned with the dark matter halos, with a misalignment angle of $\sim 35$deg. Recent work by \citet{2023arXiv230204230X} further showed that misalignment angles are likely to be mass dependent. Nevertheless, the good alignment shown in Figure~\ref{fig:wgp_bcg} seems to be in contradiction with expectations from previous studies. We found this is mainly caused by the projection effects on the observed IA of $\redmapper$ clusters, which decreases the measured cluster IA signal using member galaxy positions. If the impact on cluster IA is uncorrected, the inferred misalignment angle between BCGs and clusters are smaller than they should have been.   

\section{Summary}
\label{sec: sum}

We measured the IA of galaxy clusters by cross correlating the shapes of $\redmapper$ clusters with the LOWZ galaxie at $0.1\le z\le0.33$. We detected a positive IA signal, indicating that clusters point towards the density field. We also divide the samples into four richness samples, enabling us to study the dependence on cluster richness.  

We investigated the impact of projection effects on the measured IA of clusters using mock cluster catalogues. The inclusion of the projection effects decrease the measured IA signal by a factor of $\sim 2.5$, which is almost independent of the cluster richness. The projection effects predominantly impact the measured cluster shapes, including interlopers that are not members of the clusters and missing true members. Consequently, projection effects lead to a smaller observed misalignment angle between BCG and clusters than the underlying one.

In our study, we discovered a correlation between cluster IA and projection effects. Clusters oriented parallel to the LOS are less likely to have undetected members and more likely to have interlopers, and their projected shapes are less elliptical and exhibit weaker alignment signals. This can be attributed to their likely location within a filamentary structure along the LOS direction. Conversely, clusters oriented perpendicular to the LOS direction display a more elliptical projected shape and a stronger IA signal, they also tend to have a higher fraction of missed cluster members and a lower fraction of interlopers.

The measured IA strength, $A_{\rm IA}$, in the cluster mock with projection effects agrees well with observation. The observed $A_{\rm IA}$ in both real data and mock observe clusters barely depends on cluster richness, while a weak dependence on richness does exist if we can correctly identify the true cluster members without any contamination. 

Our work showed that IA measurements of galaxy clusters can be improved by identifying interlopers and by including the true member galaxies in the outer region, leading to a much higher signal-to-noise detection of cluster IA. High signal-to-noise detection of cluster IA is crucial for applying IA as a novel cosmological probe.
With more and more incoming spectroscopic data, we expect to suppress (or reduce) the impact of projection effects significantly. We will leave the efforts on removing projection effects for galaxy clusters to the future work. 

\section*{Acknowledgements}
We thank Teppei~Okumura, Elisa~Chisari, Ravi~K.~Sheth, Atsushi~Taruya, and Khee-Gan~Lee for enlightening discussion/comments on this work. 
This work was supported in part by World Premier Inter- national Research Center Initiative (WPI Initiative), MEXT, Japan, and JSPS KAKENHI Grant Numbers JP19H00677, JP20H05850, JP20H05855, JP20H05861, JP21H01081, and JP22K03634,
and by Basic Research Grant (Super AI) of Institute for AI and Beyond of the University of Tokyo. 
The authors thank the Yukawa Institute for Theoretical Physics at Kyoto University. Discussions during the YITP workshop YITP-W-22-16 on “New Frontiers in Cosmology with the Intrinsic Alignments of Galaxies” were useful to complete this work.
J.~Shi and T.~Kurita also thank Lorentz Center and the organizers of the hol-IA workshop: a holistic approach to galaxy intrinsic alignments held from 13 to 17 March 2023.

\section*{Data Availability}
The data underlying this article will be shared on reasonable request to the corresponding author.

\bibliographystyle{mnras}
\bibliography{ref}

\begin{thebibliography}{}
\makeatletter
\relax
\def\mn@urlcharsother{\let\do\@makeother \do\$\do\&\do\#\do\^\do\_\do\%\do\~}
\def\mn@doi{\begingroup\mn@urlcharsother \@ifnextchar [ {\mn@doi@}
  {\mn@doi@[]}}
\def\mn@doi@[#1]#2{\def\@tempa{#1}\ifx\@tempa\@empty \href
  {http://dx.doi.org/#2} {doi:#2}\else \href {http://dx.doi.org/#2} {#1}\fi
  \endgroup}
\def\mn@eprint#1#2{\mn@eprint@#1:#2::\@nil}
\def\mn@eprint@arXiv#1{\href {http://arxiv.org/abs/#1} {{\tt arXiv:#1}}}
\def\mn@eprint@dblp#1{\href {http://dblp.uni-trier.de/rec/bibtex/#1.xml}
  {dblp:#1}}
\def\mn@eprint@#1:#2:#3:#4\@nil{\def\@tempa {#1}\def\@tempb {#2}\def\@tempc
  {#3}\ifx \@tempc \@empty \let \@tempc \@tempb \let \@tempb \@tempa \fi \ifx
  \@tempb \@empty \def\@tempb {arXiv}\fi \@ifundefined
  {mn@eprint@\@tempb}{\@tempb:\@tempc}{\expandafter \expandafter \csname
  mn@eprint@\@tempb\endcsname \expandafter{\@tempc}}}

\bibitem[\protect\citeauthoryear{{Aihara} et~al.,}{{Aihara}
  et~al.}{2011}]{2011ApJS..193...29A}
{Aihara} H.,  et~al., 2011, \mn@doi [\apjs] {10.1088/0067-0049/193/2/29}, \href
  {https://ui.adsabs.harvard.edu/abs/2011ApJS..193...29A} {193, 29}

\bibitem[\protect\citeauthoryear{{Akitsu}, {Li}  \& {Okumura}}{{Akitsu}
  et~al.}{2021}]{2021JCAP...04..041A}
{Akitsu} K.,  {Li} Y.,   {Okumura} T.,  2021, \mn@doi [\jcap]
  {10.1088/1475-7516/2021/04/041}, \href
  {https://ui.adsabs.harvard.edu/abs/2021JCAP...04..041A} {2021, 041}

\bibitem[\protect\citeauthoryear{{Anderson} et~al.,}{{Anderson}
  et~al.}{2012}]{2012MNRAS.427.3435A}
{Anderson} L.,  et~al., 2012, \mn@doi [\mnras]
  {10.1111/j.1365-2966.2012.22066.x}, \href
  {https://ui.adsabs.harvard.edu/abs/2012MNRAS.427.3435A} {427, 3435}

\bibitem[\protect\citeauthoryear{{Behroozi}, {Wechsler}  \& {Wu}}{{Behroozi}
  et~al.}{2013}]{2013ApJ...762..109B}
{Behroozi} P.~S.,  {Wechsler} R.~H.,   {Wu} H.-Y.,  2013, \mn@doi [\apj]
  {10.1088/0004-637X/762/2/109}, \href
  {https://ui.adsabs.harvard.edu/abs/2013ApJ...762..109B} {762, 109}

\bibitem[\protect\citeauthoryear{{Bernstein} \& {Jarvis}}{{Bernstein} \&
  {Jarvis}}{2002}]{2002AJ....123..583B}
{Bernstein} G.~M.,  {Jarvis} M.,  2002, \mn@doi [\aj] {10.1086/338085}, \href
  {https://ui.adsabs.harvard.edu/abs/2002AJ....123..583B} {123, 583}

\bibitem[\protect\citeauthoryear{{Bridle} \& {King}}{{Bridle} \&
  {King}}{2007}]{2007NJPh....9..444B}
{Bridle} S.,  {King} L.,  2007, \mn@doi [New Journal of Physics]
  {10.1088/1367-2630/9/12/444}, \href
  {https://ui.adsabs.harvard.edu/abs/2007NJPh....9..444B} {9, 444}

\bibitem[\protect\citeauthoryear{{Catelan}, {Kamionkowski}  \&
  {Blandford}}{{Catelan} et~al.}{2001}]{2001MNRAS.320L...7C}
{Catelan} P.,  {Kamionkowski} M.,   {Blandford} R.~D.,  2001, \mn@doi [\mnras]
  {10.1046/j.1365-8711.2001.04105.x}, \href
  {https://ui.adsabs.harvard.edu/abs/2001MNRAS.320L...7C} {320, L7}

\bibitem[\protect\citeauthoryear{{Cohn}, {Evrard}, {White}, {Croton}  \&
  {Ellingson}}{{Cohn} et~al.}{2007}]{2007MNRAS.382.1738C}
{Cohn} J.~D.,  {Evrard} A.~E.,  {White} M.,  {Croton} D.,   {Ellingson} E.,
  2007, \mn@doi [\mnras] {10.1111/j.1365-2966.2007.12479.x}, \href
  {https://ui.adsabs.harvard.edu/abs/2007MNRAS.382.1738C} {382, 1738}

\bibitem[\protect\citeauthoryear{{Costanzi} et~al.,}{{Costanzi}
  et~al.}{2019}]{2019MNRAS.482..490C}
{Costanzi} M.,  et~al., 2019, \mn@doi [\mnras] {10.1093/mnras/sty2665}, \href
  {https://ui.adsabs.harvard.edu/abs/2019MNRAS.482..490C} {482, 490}

\bibitem[\protect\citeauthoryear{{Evans} \& {Bridle}}{{Evans} \&
  {Bridle}}{2009}]{2009ApJ...695.1446E}
{Evans} A. K.~D.,  {Bridle} S.,  2009, \mn@doi [\apj]
  {10.1088/0004-637X/695/2/1446}, \href
  {https://ui.adsabs.harvard.edu/abs/2009ApJ...695.1446E} {695, 1446}

\bibitem[\protect\citeauthoryear{{Gonzalez}, {Hoffmann}, {Gazta{\~n}aga},
  {Garc{\'\i}a Lambas}, {Fosalba}, {Crocce}, {Castander}  \&
  {Makler}}{{Gonzalez} et~al.}{2022}]{2022MNRAS.517.4827G}
{Gonzalez} E.~J.,  {Hoffmann} K.,  {Gazta{\~n}aga} E.,  {Garc{\'\i}a Lambas}
  D.~R.,  {Fosalba} P.,  {Crocce} M.,  {Castander} F.~J.,   {Makler} M.,  2022,
  \mn@doi [\mnras] {10.1093/mnras/stac3038}, \href
  {https://ui.adsabs.harvard.edu/abs/2022MNRAS.517.4827G} {517, 4827}

\bibitem[\protect\citeauthoryear{{Hamilton}}{{Hamilton}}{2015}]{2015ascl.soft12017H}
{Hamilton} A. J.~S.,  2015, {FFTLog: Fast Fourier or Hankel transform},
  Astrophysics Source Code Library, record ascl:1512.017 (\mn@eprint {ascl}
  {1512.017})

\bibitem[\protect\citeauthoryear{{Hinshaw} et~al.,}{{Hinshaw}
  et~al.}{2013}]{2013ApJS..208...19H}
{Hinshaw} G.,  et~al., 2013, \mn@doi [\apjs] {10.1088/0067-0049/208/2/19},
  \href {https://ui.adsabs.harvard.edu/abs/2013ApJS..208...19H} {208, 19}

\bibitem[\protect\citeauthoryear{{Hirata} \& {Seljak}}{{Hirata} \&
  {Seljak}}{2004}]{2004PhRvD..70f3526H}
{Hirata} C.~M.,  {Seljak} U.,  2004, \mn@doi [\prd]
  {10.1103/PhysRevD.70.063526}, \href
  {https://ui.adsabs.harvard.edu/abs/2004PhRvD..70f3526H} {70, 063526}

\bibitem[\protect\citeauthoryear{{Jarvis}, {Bernstein}  \& {Jain}}{{Jarvis}
  et~al.}{2004}]{2004MNRAS.352..338J}
{Jarvis} M.,  {Bernstein} G.,   {Jain} B.,  2004, \mn@doi [\mnras]
  {10.1111/j.1365-2966.2004.07926.x}, \href
  {https://ui.adsabs.harvard.edu/abs/2004MNRAS.352..338J} {352, 338}

\bibitem[\protect\citeauthoryear{{Joachimi}, {Mandelbaum}, {Abdalla}  \&
  {Bridle}}{{Joachimi} et~al.}{2011}]{2011A&A...527A..26J}
{Joachimi} B.,  {Mandelbaum} R.,  {Abdalla} F.~B.,   {Bridle} S.~L.,  2011,
  \mn@doi [\aap] {10.1051/0004-6361/201015621}, \href
  {https://ui.adsabs.harvard.edu/abs/2011A&A...527A..26J} {527, A26}

\bibitem[\protect\citeauthoryear{{Joachimi} et~al.,}{{Joachimi}
  et~al.}{2015}]{2015SSRv..193....1J}
{Joachimi} B.,  et~al., 2015, \mn@doi [\ssr] {10.1007/s11214-015-0177-4}, \href
  {https://ui.adsabs.harvard.edu/abs/2015SSRv..193....1J} {193, 1}

\bibitem[\protect\citeauthoryear{{Kaiser}}{{Kaiser}}{1987}]{Kaiser1987:RSD}
{Kaiser} N.,  1987, \mn@doi [\mnras] {10.1093/mnras/227.1.1}, \href
  {https://ui.adsabs.harvard.edu/abs/1987MNRAS.227....1K} {227, 1}

\bibitem[\protect\citeauthoryear{{Kiessling} et~al.,}{{Kiessling}
  et~al.}{2015}]{2015SSRv..193...67K}
{Kiessling} A.,  et~al., 2015, \mn@doi [\ssr] {10.1007/s11214-015-0203-6},
  \href {https://ui.adsabs.harvard.edu/abs/2015SSRv..193...67K} {193, 67}

\bibitem[\protect\citeauthoryear{{Kirk} et~al.,}{{Kirk}
  et~al.}{2015}]{2015SSRv..193..139K}
{Kirk} D.,  et~al., 2015, \mn@doi [\ssr] {10.1007/s11214-015-0213-4}, \href
  {https://ui.adsabs.harvard.edu/abs/2015SSRv..193..139K} {193, 139}

\bibitem[\protect\citeauthoryear{{Kurita} \& {Takada}}{{Kurita} \&
  {Takada}}{2022}]{2022PhRvD.105l3501K}
{Kurita} T.,  {Takada} M.,  2022, \mn@doi [\prd] {10.1103/PhysRevD.105.123501},
  \href {https://ui.adsabs.harvard.edu/abs/2022PhRvD.105l3501K} {105, 123501}

\bibitem[\protect\citeauthoryear{{Kurita} \& {Takada}}{{Kurita} \&
  {Takada}}{2023}]{2023arXiv230202925K}
{Kurita} T.,  {Takada} M.,  2023, \mn@doi [arXiv e-prints]
  {10.48550/arXiv.2302.02925}, \href
  {https://ui.adsabs.harvard.edu/abs/2023arXiv230202925K} {p. arXiv:2302.02925}

\bibitem[\protect\citeauthoryear{{Kurita}, {Takada}, {Nishimichi}, {Takahashi},
  {Osato}  \& {Kobayashi}}{{Kurita} et~al.}{2021}]{2021MNRAS.501..833K}
{Kurita} T.,  {Takada} M.,  {Nishimichi} T.,  {Takahashi} R.,  {Osato} K.,
  {Kobayashi} Y.,  2021, \mn@doi [\mnras] {10.1093/mnras/staa3625}, \href
  {https://ui.adsabs.harvard.edu/abs/2021MNRAS.501..833K} {501, 833}

\bibitem[\protect\citeauthoryear{{Landy} \& {Szalay}}{{Landy} \&
  {Szalay}}{1993}]{1993ApJ...412...64L}
{Landy} S.~D.,  {Szalay} A.~S.,  1993, \mn@doi [\apj] {10.1086/172900}, \href
  {https://ui.adsabs.harvard.edu/abs/1993ApJ...412...64L} {412, 64}

\bibitem[\protect\citeauthoryear{{Mandelbaum} et~al.,}{{Mandelbaum}
  et~al.}{2011}]{2011MNRAS.410..844M}
{Mandelbaum} R.,  et~al., 2011, \mn@doi [\mnras]
  {10.1111/j.1365-2966.2010.17485.x}, \href
  {https://ui.adsabs.harvard.edu/abs/2011MNRAS.410..844M} {410, 844}

\bibitem[\protect\citeauthoryear{{Navarro}, {Frenk}  \& {White}}{{Navarro}
  et~al.}{1997}]{1997ApJ...490..493N}
{Navarro} J.~F.,  {Frenk} C.~S.,   {White} S. D.~M.,  1997, \mn@doi [\apj]
  {10.1086/304888}, \href
  {https://ui.adsabs.harvard.edu/abs/1997ApJ...490..493N} {490, 493}

\bibitem[\protect\citeauthoryear{{Nishimichi} et~al.,}{{Nishimichi}
  et~al.}{2019}]{2019ApJ...884...29N}
{Nishimichi} T.,  et~al., 2019, \mn@doi [\apj] {10.3847/1538-4357/ab3719},
  \href {https://ui.adsabs.harvard.edu/abs/2019ApJ...884...29N} {884, 29}

\bibitem[\protect\citeauthoryear{{Norberg}, {Baugh}, {Gazta{\~n}aga}  \&
  {Croton}}{{Norberg} et~al.}{2009}]{2009MNRAS.396...19N}
{Norberg} P.,  {Baugh} C.~M.,  {Gazta{\~n}aga} E.,   {Croton} D.~J.,  2009,
  \mn@doi [\mnras] {10.1111/j.1365-2966.2009.14389.x}, \href
  {https://ui.adsabs.harvard.edu/abs/2009MNRAS.396...19N} {396, 19}

\bibitem[\protect\citeauthoryear{{Oguri}, {Takada}, {Okabe}  \&
  {Smith}}{{Oguri} et~al.}{2010}]{2010MNRAS.405.2215O}
{Oguri} M.,  {Takada} M.,  {Okabe} N.,   {Smith} G.~P.,  2010, \mn@doi [\mnras]
  {10.1111/j.1365-2966.2010.16622.x}, \href
  {https://ui.adsabs.harvard.edu/abs/2010MNRAS.405.2215O} {405, 2215}

\bibitem[\protect\citeauthoryear{{Okumura}, {Jing}  \& {Li}}{{Okumura}
  et~al.}{2009}]{2009ApJ...694..214O}
{Okumura} T.,  {Jing} Y.~P.,   {Li} C.,  2009, \mn@doi [\apj]
  {10.1088/0004-637X/694/1/214}, \href
  {https://ui.adsabs.harvard.edu/abs/2009ApJ...694..214O} {694, 214}

\bibitem[\protect\citeauthoryear{{Osato}, {Nishimichi}, {Oguri}, {Takada}  \&
  {Okumura}}{{Osato} et~al.}{2018}]{2018MNRAS.477.2141O}
{Osato} K.,  {Nishimichi} T.,  {Oguri} M.,  {Takada} M.,   {Okumura} T.,  2018,
  \mn@doi [\mnras] {10.1093/mnras/sty762}, \href
  {https://ui.adsabs.harvard.edu/abs/2018MNRAS.477.2141O} {477, 2141}

\bibitem[\protect\citeauthoryear{{Park}, {Sunayama}, {Takada}, {Kobayashi},
  {Miyatake}, {More}, {Nishimichi}  \& {Sugiyama}}{{Park}
  et~al.}{2023}]{2023MNRAS.518.5171P}
{Park} Y.,  {Sunayama} T.,  {Takada} M.,  {Kobayashi} Y.,  {Miyatake} H.,
  {More} S.,  {Nishimichi} T.,   {Sugiyama} S.,  2023, \mn@doi [\mnras]
  {10.1093/mnras/stac3410}, \href
  {https://ui.adsabs.harvard.edu/abs/2023MNRAS.518.5171P} {518, 5171}

\bibitem[\protect\citeauthoryear{{Planck Collaboration} et~al.,}{{Planck
  Collaboration} et~al.}{2016}]{2016A&A...594A..13P}
{Planck Collaboration} et~al., 2016, \mn@doi [\aap]
  {10.1051/0004-6361/201525830}, \href
  {https://ui.adsabs.harvard.edu/abs/2016A&A...594A..13P} {594, A13}

\bibitem[\protect\citeauthoryear{{Reyes}, {Mandelbaum}, {Gunn}, {Nakajima},
  {Seljak}  \& {Hirata}}{{Reyes} et~al.}{2012}]{2012MNRAS.425.2610R}
{Reyes} R.,  {Mandelbaum} R.,  {Gunn} J.~E.,  {Nakajima} R.,  {Seljak} U.,
  {Hirata} C.~M.,  2012, \mn@doi [\mnras] {10.1111/j.1365-2966.2012.21472.x},
  \href {https://ui.adsabs.harvard.edu/abs/2012MNRAS.425.2610R} {425, 2610}

\bibitem[\protect\citeauthoryear{{Rozo} \& {Rykoff}}{{Rozo} \&
  {Rykoff}}{2014}]{2014ApJ...783...80R}
{Rozo} E.,  {Rykoff} E.~S.,  2014, \mn@doi [\apj] {10.1088/0004-637X/783/2/80},
  \href {https://ui.adsabs.harvard.edu/abs/2014ApJ...783...80R} {783, 80}

\bibitem[\protect\citeauthoryear{{Rykoff} et~al.,}{{Rykoff}
  et~al.}{2014}]{2014ApJ...785..104R}
{Rykoff} E.~S.,  et~al., 2014, \mn@doi [\apj] {10.1088/0004-637X/785/2/104},
  \href {https://ui.adsabs.harvard.edu/abs/2014ApJ...785..104R} {785, 104}

\bibitem[\protect\citeauthoryear{{Rykoff} et~al.,}{{Rykoff}
  et~al.}{2016}]{2016ApJS..224....1R}
{Rykoff} E.~S.,  et~al., 2016, \mn@doi [\apjs] {10.3847/0067-0049/224/1/1},
  \href {https://ui.adsabs.harvard.edu/abs/2016ApJS..224....1R} {224, 1}

\bibitem[\protect\citeauthoryear{{Schmidt}, {Pajer}  \&
  {Zaldarriaga}}{{Schmidt} et~al.}{2014}]{2014PhRvD..89h3507S}
{Schmidt} F.,  {Pajer} E.,   {Zaldarriaga} M.,  2014, \mn@doi [\prd]
  {10.1103/PhysRevD.89.083507}, \href
  {https://ui.adsabs.harvard.edu/abs/2014PhRvD..89h3507S} {89, 083507}

\bibitem[\protect\citeauthoryear{{Shin}, {Clampitt}, {Jain}, {Bernstein},
  {Neil}, {Rozo}  \& {Rykoff}}{{Shin} et~al.}{2018}]{2018MNRAS.475.2421S}
{Shin} T.-h.,  {Clampitt} J.,  {Jain} B.,  {Bernstein} G.,  {Neil} A.,  {Rozo}
  E.,   {Rykoff} E.,  2018, \mn@doi [\mnras] {10.1093/mnras/stx3366}, \href
  {https://ui.adsabs.harvard.edu/abs/2018MNRAS.475.2421S} {475, 2421}

\bibitem[\protect\citeauthoryear{{Simet}, {McClintock}, {Mandelbaum}, {Rozo},
  {Rykoff}, {Sheldon}  \& {Wechsler}}{{Simet}
  et~al.}{2017}]{2017MNRAS.466.3103S}
{Simet} M.,  {McClintock} T.,  {Mandelbaum} R.,  {Rozo} E.,  {Rykoff} E.,
  {Sheldon} E.,   {Wechsler} R.~H.,  2017, \mn@doi [\mnras]
  {10.1093/mnras/stw3250}, \href
  {https://ui.adsabs.harvard.edu/abs/2017MNRAS.466.3103S} {466, 3103}

\bibitem[\protect\citeauthoryear{{Singh}, {Mandelbaum}  \& {More}}{{Singh}
  et~al.}{2015}]{2015MNRAS.450.2195S}
{Singh} S.,  {Mandelbaum} R.,   {More} S.,  2015, \mn@doi [\mnras]
  {10.1093/mnras/stv778}, \href
  {https://ui.adsabs.harvard.edu/abs/2015MNRAS.450.2195S} {450, 2195}

\bibitem[\protect\citeauthoryear{{Smargon}, {Mandelbaum}, {Bahcall}  \&
  {Niederste-Ostholt}}{{Smargon} et~al.}{2012}]{2012MNRAS.423..856S}
{Smargon} A.,  {Mandelbaum} R.,  {Bahcall} N.,   {Niederste-Ostholt} M.,  2012,
  \mn@doi [\mnras] {10.1111/j.1365-2966.2012.20923.x}, \href
  {https://ui.adsabs.harvard.edu/abs/2012MNRAS.423..856S} {423, 856}

\bibitem[\protect\citeauthoryear{{Sunayama}}{{Sunayama}}{2022}]{2022arXiv220503233S}
{Sunayama} T.,  2022, \mn@doi [arXiv e-prints] {10.48550/arXiv.2205.03233},
  \href {https://ui.adsabs.harvard.edu/abs/2022arXiv220503233S} {p.
  arXiv:2205.03233}

\bibitem[\protect\citeauthoryear{{Sunayama} \& {More}}{{Sunayama} \&
  {More}}{2019}]{2019MNRAS.490.4945S}
{Sunayama} T.,  {More} S.,  2019, \mn@doi [\mnras] {10.1093/mnras/stz2832},
  \href {https://ui.adsabs.harvard.edu/abs/2019MNRAS.490.4945S} {490, 4945}

\bibitem[\protect\citeauthoryear{{Sunayama} et~al.,}{{Sunayama}
  et~al.}{2020}]{2020MNRAS.496.4468S}
{Sunayama} T.,  et~al., 2020, \mn@doi [\mnras] {10.1093/mnras/staa1646}, \href
  {https://ui.adsabs.harvard.edu/abs/2020MNRAS.496.4468S} {496, 4468}

\bibitem[\protect\citeauthoryear{{Takahashi}, {Sato}, {Nishimichi}, {Taruya}
  \& {Oguri}}{{Takahashi} et~al.}{2012}]{2012ApJ...761..152T}
{Takahashi} R.,  {Sato} M.,  {Nishimichi} T.,  {Taruya} A.,   {Oguri} M.,
  2012, \mn@doi [\apj] {10.1088/0004-637X/761/2/152}, \href
  {https://ui.adsabs.harvard.edu/abs/2012ApJ...761..152T} {761, 152}

\bibitem[\protect\citeauthoryear{{To} et~al.,}{{To}
  et~al.}{2021}]{2021PhRvL.126n1301T}
{To} C.,  et~al., 2021, \mn@doi [\prl] {10.1103/PhysRevLett.126.141301}, \href
  {https://ui.adsabs.harvard.edu/abs/2021PhRvL.126n1301T} {126, 141301}

\bibitem[\protect\citeauthoryear{{Troxel} \& {Ishak}}{{Troxel} \&
  {Ishak}}{2015}]{2015PhR...558....1T}
{Troxel} M.~A.,  {Ishak} M.,  2015, \mn@doi [\physrep]
  {10.1016/j.physrep.2014.11.001}, \href
  {https://ui.adsabs.harvard.edu/abs/2015PhR...558....1T} {558, 1}

\bibitem[\protect\citeauthoryear{{Weinberg}, {Mortonson}, {Eisenstein},
  {Hirata}, {Riess}  \& {Rozo}}{{Weinberg} et~al.}{2013}]{2013PhR...530...87W}
{Weinberg} D.~H.,  {Mortonson} M.~J.,  {Eisenstein} D.~J.,  {Hirata} C.,
  {Riess} A.~G.,   {Rozo} E.,  2013, \mn@doi [\physrep]
  {10.1016/j.physrep.2013.05.001}, \href
  {https://ui.adsabs.harvard.edu/abs/2013PhR...530...87W} {530, 87}

\bibitem[\protect\citeauthoryear{{Xu}, {Jing}  \& {Gao}}{{Xu}
  et~al.}{2023}]{2023arXiv230204230X}
{Xu} K.,  {Jing} Y.~P.,   {Gao} H.,  2023, \mn@doi [arXiv e-prints]
  {10.48550/arXiv.2302.04230}, \href
  {https://ui.adsabs.harvard.edu/abs/2023arXiv230204230X} {p. arXiv:2302.04230}

\bibitem[\protect\citeauthoryear{{Yao}, {Shan}, {Zhang}, {Kneib}  \&
  {Jullo}}{{Yao} et~al.}{2020}]{2020ApJ...904..135Y}
{Yao} J.,  {Shan} H.,  {Zhang} P.,  {Kneib} J.-P.,   {Jullo} E.,  2020, \mn@doi
  [\apj] {10.3847/1538-4357/abc175}, \href
  {https://ui.adsabs.harvard.edu/abs/2020ApJ...904..135Y} {904, 135}

\bibitem[\protect\citeauthoryear{{Zheng} et~al.,}{{Zheng}
  et~al.}{2005}]{2005ApJ...633..791Z}
{Zheng} Z.,  et~al., 2005, \mn@doi [\apj] {10.1086/466510}, \href
  {https://ui.adsabs.harvard.edu/abs/2005ApJ...633..791Z} {633, 791}

\bibitem[\protect\citeauthoryear{{van Haarlem}, {Frenk}  \& {White}}{{van
  Haarlem} et~al.}{1997}]{1997MNRAS.287..817V}
{van Haarlem} M.~P.,  {Frenk} C.~S.,   {White} S.~D.~M.,  1997, \mn@doi
  [\mnras] {10.1093/mnras/287.4.817}, \href
  {https://ui.adsabs.harvard.edu/abs/1997MNRAS.287..817V} {287, 817}

\bibitem[\protect\citeauthoryear{{van Uitert} \& {Joachimi}}{{van Uitert} \&
  {Joachimi}}{2017}]{2017MNRAS.468.4502V}
{van Uitert} E.,  {Joachimi} B.,  2017, \mn@doi [\mnras]
  {10.1093/mnras/stx756}, \href
  {https://ui.adsabs.harvard.edu/abs/2017MNRAS.468.4502V} {468, 4502}

\bibitem[\protect\citeauthoryear{{van den Bosch}, {More}, {Cacciato}, {Mo}  \&
  {Yang}}{{van den Bosch} et~al.}{2013}]{2013MNRAS.430..725V}
{van den Bosch} F.~C.,  {More} S.,  {Cacciato} M.,  {Mo} H.,   {Yang} X.,
  2013, \mn@doi [\mnras] {10.1093/mnras/sts006}, \href
  {https://ui.adsabs.harvard.edu/abs/2013MNRAS.430..725V} {430, 725}

\makeatother
\end{thebibliography}

\newpage
\appendix
\counterwithin{figure}{section}

\section{Numerical Implementation of Two-point Correlation Function}
\label{app:2pt}

We here review the three-dimensional two-point statistics of shear. 
The goal of this section is to derive Eq.~\eqref{eq:IAxi_multipoles_def} in the main text. 

\subsection{Two-point Statistics}

We assume the distant-observer (plane-parallel) approximation throughout this section. 
The shear of a cluster at a position $\bx$ is given by 
\begin{align}
    \gamma(\bx) = \gamma_1(\bx) + i\gamma_2(\bx). 
    \label{eq:def_gamma}
\end{align}
This is a spin-2 quantity on the sky plane perpendicular to the line-of-sight direction (LOS). 
To obtain the coordinate-independent shear for the two-point correlation function, we define the rotated shear with the radial and cross components towards the other galaxy in a pair at a position $\bx'$ as
\begin{align}
    \gamma_{+, \times}(\bx;\bx') \equiv \gamma(\bx) e^{-2i\phi_{\br}}, 
    \label{eq:def_gamma_rot}
\end{align}
where $\br \equiv \bx-\bx'$ is the separation vector and $\phi_{\br}$ is the angle measured from the first coordinate axis to the projected separation vector $\br_\mathrm{p}$ on the sky plane. 
The two-point cross-correlation function of the galaxy density and shear is defined by 
\begin{align}
    \xi_{\rmg \gamma}(\br) 
    \equiv \avrg{\gamma_{+, \times}(\bx;\bx') \delg(\bx')} 
    = \avrg{\gamma(\bx) \delg(\bx')} e^{-2i\phi_{\br}}, 
    \label{eq:def_2pcf_cross}
\end{align}
where the radial and cross components correspond to the real and imaginary parts, $\xi_\mathrm{g+}=\Re\, \xi_{\rmg\gamma}$ and $\xi_\mathrm{g\times}=\Im\, \xi_{\rmg\gamma}$, respectively.

In Fourier space, we start with the Fourier transform of Eq.~\eqref{eq:def_gamma}: 
\begin{align}
    \gamma(\bk) = \gamma_1(\bk) + i\gamma_2(\bk). 
    \label{eq:def_gammaF}
\end{align}
As in the case of configuration space, we define the coordinate-independent quantities in Fourier space, called $E/B$ modes, with a similar rotation as
\begin{align}
    E(\bk)+iB(\bk) \equiv \gamma(\bk) e^{-2i\phi_{\bk}}, 
    \label{eq:def_EB}
\end{align}
where $\phi_{\bk}$ is the angle measured from the first coordinate axis to the wave vector on the sky plane. 
The cross power spectrum of the galaxy density and shear is thus given by
\begin{align}
    (2\pi)^3 \delta_\rmD (\bk+\bk') P_{\rmg \gamma}(\bk)
    &\equiv \avrg{[E(\bk)+iB(\bk)] \delg(\bk')} \nonumber\\
    &= \avrg{\gamma(\bk) \delg(\bk')} e^{-2i\phi_{\bk}}, 
    \label{eq:def_ps_cross}
\end{align}
where the $E$- and $B$-mode spectra correspond to $P_\mathrm{gE} = \Re\, P_{\rmg \gamma}$ and $P_\mathrm{gB} = \Im\, P_{\rmg \gamma}$, respectively.

From Eqs.~\eqref{eq:def_2pcf_cross} and \eqref{eq:def_ps_cross}, we obtain the relation between the correlation function and the power spectrum, 
\begin{align}
    \xi_{\rmg\gamma}(\br) 
    = \int \frac{\rmd \bk}{(2\pi)^3}\, 
    P_{\rmg \gamma}(\bk) e^{2i(\phi_{\bk} - \phi_{\br})} e^{i\bk\cdot\br}. 
    \label{eq:relation_2pcf_ps_cross}
\end{align}
These statistics are anisotropic with respect to the LOS due to the RSD and the projection of galaxy shape to the sky plane: $\xi_{\rmg\gamma}(\br) = \xi_{\rmg\gamma}(r_\mathrm{p}, r_\parallel)$ and $P_{\rmg \gamma}(\bk) = P_{\rmg \gamma}(k_\mathrm{p}, k_\parallel)$, respectively.

The projected correlation function is defined by the integral of the correlation function over the LOS: 
\begin{align}
    w_{\rmg\gamma}(r_\mathrm{p}; \Pi_{\rm max}, z) 
    = \int_{-\Pi_{\rm max}}^{\Pi_{\rm max}} \rmd r_\parallel\, \xi_{\rmg\gamma}(r_\mathrm{p}, r_\parallel, z), 
    \label{eq:def_proj_cf_cross}
\end{align}
where $\Pi_{\rm max}$ is the projection length of the LOS direction for which an observer needs to specify; as our default choice, we adopt $\Pi_{\rm max}=100~h^{-1}{\rm Mpc}$.
This expression corresponds to the projected correlation at a single, representative redshift.
If we take into account the redshift dependence, we can follow the method in \citet{2015MNRAS.450.2195S} as
\begin{align}
    w_{\rmg\gamma}(r_\mathrm{p}; \Pi_{\rm max}, \bar{z}) 
    \equiv \int \rmd z\, W(z) w_{\rmg\gamma}(r_\mathrm{p}; \Pi_{\rm max}, z), 
\end{align}
where $W(z)$ is the redshift distribution of the galaxy density and shape tracers, defined as
\begin{align}
W(z)=\frac{p_g(z)p_\gamma(z)}{\chi^2\mathrm{d\chi}/\mathrm{d}z}
\left[\int\!\mathrm{d}z~
\frac{p_g(z)p_\gamma(z)}{\chi^2\mathrm{d\chi}/\mathrm{d}z}
\right]^{-1}.
\end{align}

\subsection{Expression with Spherical Bessel Function}
To numerically evaluate the correlation function $\xi_{\rmg\gamma}$, one has to compute the transform in Eq.~\eqref{eq:relation_2pcf_ps_cross} from the input model $P_{\rmg \gamma}$. 
The standard method is to use the isotropy around the LOS on the sky plane and integrate it in the cylindrical coordinates \cite[e.g.][]{2015MNRAS.450.2195S}. 
In this work, we employ the spherical coordinates and use an alternative expression with the spherical Bessel function derived in \cite{2022PhRvD.105l3501K}. 
We here briefly review the derivation. 

First, we decompose the model power spectrum into the multipoles of the associated Legendre polynomials with $m=2$, $\mathcal{L}_\ell^2$, as
\begin{align}
    P_{\rmg \gamma}(k_\mathrm{p}, k_\parallel) = 
    P_{\rmg \gamma}(k, \mu_{\bk}) = 
    \sum_{\ell \geq 2} P_{\rmg \gamma}^{(\ell)}(k) \mathcal{L}_\ell^2(\mu_{\bk}), 
    \label{eq:def_ps_multipole}
\end{align}
where $\mu_{\bk} \equiv \hbk \cdot \hbn = k_\parallel/k$ is the cosine between the wave vector and the LOS $\hbn$. 
Note $\mathcal{L}_2^2(x)=3(1-x^2)$, $\mathcal{L}_4^2(x)=15(1-x^2)(7x^2-1)/2$, and so forth.
Substituting Eq.~\eqref{eq:def_ps_multipole} into Eq.~\eqref{eq:relation_2pcf_ps_cross} and employing the spherical coordinates, we have 
\begin{align}
    \xi_{\rmg\gamma}(\br) 
    = \sum_{\ell \geq 2} 
    \int \frac{k^2\rmd k}{2\pi^2} \, P_{\rmg \gamma}^{(\ell)}(k)
    \int \frac{\rmd \Omega_{\bk}}{4\pi} \, \mathcal{L}_\ell^2(\mu_{\bk})
    e^{2i(\phi_{\bk} - \phi_{\br})} e^{i\bk\cdot\br}. 
\end{align}
Recalling the definition of the spherical harmonics, 
\begin{align}
    Y_\ell^m(\hbk) = N_\ell^m \mathcal{L}_\ell^m(\mu_{\bk}) e^{im\phi_{\bk}},
\end{align}
with $N_\ell^m$ being the normalization factor
\begin{align}
    N_\ell^m = \sqrt{\frac{(2\ell+1)}{4\pi} \frac{(\ell-m)!}{(\ell+m)!}}, 
\end{align}
and using the plane-wave expansion
\begin{align}
    e^{i\bk\cdot\br} = 4\pi \sum_{\ell,m} i^\ell j_\ell(kr) Y_\ell^{m*}(\hbk) Y_\ell^{m}(\hbr),
\end{align}
we carry out the angle average of the wave vector $\hbk$ as
\begin{align}
    &\quad\sum_{\ell \geq 2} \int \frac{k^2\rmd k}{2\pi^2} \, P_{\rmg \gamma}^{(\ell)}(k) 
    \int \frac{\rmd \Omega_{\bk}}{4\pi} \, \mathcal{L}_\ell^2(\mu_{\bk})
    e^{2i(\phi_{\bk} - \phi_{\br})} e^{i\bk\cdot\br} \nonumber\\
    &= \sum_{\ell \geq 2} \int \frac{k^2\rmd k}{2\pi^2} \, P_{\rmg \gamma}^{(\ell)}(k) 
    \int \frac{\rmd \Omega_{\bk}}{4\pi} \, (N_\ell^2)^{-1} Y_\ell^2(\hbk) \nonumber\\
    &\quad\quad \times 4\pi \sum_{\ell',m'} i^{\ell'} j_{\ell'}(kr) Y_{\ell'}^{m'*}(\hbk) Y_{\ell'}^{m'}(\hbr) e^{-2i\phi_{\br}} \nonumber\\
    &= \sum_{\ell \geq 2} i^{\ell} \int \frac{k^2\rmd k}{2\pi^2} \, P_{\rmg \gamma}^{(\ell)}(k) 
    j_{\ell}(kr) (N_\ell^2)^{-1} Y_{\ell}^{2}(\hbr) e^{-2i\phi_{\br}} \nonumber\\
    &= \sum_{\ell \geq 2} \left[i^{\ell} \int \frac{k^2\rmd k}{2\pi^2} \, P_{\rmg \gamma}^{(\ell)}(k) 
    j_{\ell}(kr)\right] \mathcal{L}_\ell^2(\mu_{\br}). 
\end{align}
In the second equation, we have used the orthogonality
\begin{align}
    \int \rmd \Omega_{\bk} \, Y_\ell^m(\hbk) Y_{\ell'}^{m'*}(\hbk) = \delta_{\ell\ell'}\delta_{mm'}. 
\end{align}
By comparing this result and the multipoles of the correlation function defined by
\begin{align}
    \xi_{\rmg \gamma}(r_\mathrm{p}, r_\parallel) = \sum_{\ell \geq 2} \xi_{\rmg \gamma}^{(\ell)}(r) \mathcal{L}_\ell^2(\mu_{\br}), 
    \label{eq:def_2pcf_multipole}
\end{align}
where $\mu_{\br} \equiv \hbr \cdot \hbn = r_\parallel/r$, we obtain the expression of 
the multipoles
\begin{align}
    \xi_{\rmg \gamma}^{(\ell)}(r) = i^{\ell} \int \frac{k^2\rmd k}{2\pi^2} \, P_{\rmg \gamma}^{(\ell)}(k) 
    j_{\ell}(kr), 
    \label{eq:hankel_ps2cf}
\end{align}
which can be computed by the use of FFTlog algorithm \citep{2015ascl.soft12017H}.

Let us consider the linear model, i.e. linear alignment model \citep{2004PhRvD..70f3526H} with Kaiser formula \citep{Kaiser1987:RSD}, as an example.
The model power spectrum is given by
\begin{align}
    P_{\rmg \gamma}(k, \mu_{\bk}) = \frac{1-\mu_{\bk}^2}{2}(1+\beta\mu_{\bk}^2) b_gb_K P_\mathrm{mm}(k),
\end{align}
where $\beta \equiv f/b_g$, $P_\mathrm{mm}$ is the linear matter power spectrum, $b_g$ and $b_K \equiv -2A_\mathrm{IA}C_1\rho_\mathrm{crit}\Omega_\rmm/\bar{D}$ are the linear bias of the density sample and shape bias, respectively. 
The multipole coefficients of the associated Legendre polynomials then become
\begin{align}
    P_{\rmg \gamma}^{(2)}(k) &= \frac{1}{6} \left(1+\frac{\beta}{7} \right) b_g b_KP_\mathrm{mm}(k),~\\
    P_{\rmg \gamma}^{(4)}(k) &= \frac{1}{105} \beta b_g b_K P_\mathrm{mm}(k),
    \label{eq:IA_multipoles_def}
\end{align}
and zero otherwise. 
Plugging these into Eq.~\eqref{eq:hankel_ps2cf}, we obtain the multipoles of correlation function with the Hankel transforms of the input matter power spectrum:
\begin{align}
    \xi_{\rmg \gamma}^{(2)}(r) &= \frac{1}{6} \left(1+\frac{\beta}{7} \right) b_g b_K \xi_\mathrm{mm}^{(2)}(r),~\\
    \xi_{\rmg \gamma}^{(4)}(r) &= \frac{1}{105} \beta b_g b_K \xi_\mathrm{mm}^{(4)}(r),
\end{align}
where we have defined the multipoles of matter correlation function:
\begin{align}
    \xi_\mathrm{mm}^{(\ell)}(r) \equiv i^{\ell} \int \frac{k^2\rmd k}{2\pi^2} \, P_\mathrm{mm}(k) 
    j_{\ell}(kr), 
\end{align}
Once we prepare these multipoles, we can obtain the projected correlation function by integrating over the LOS as in Eq.~\eqref{eq:def_proj_cf_cross},
\begin{align}
    w_{\rmg\gamma}(r_\mathrm{p}; \Pi, z) 
    &= \int_{-\Pi_{\rm max}}^{\Pi_{\rm max}} \rmd r_\parallel\, \xi_{\rmg\gamma}(r_\mathrm{p}, r_\parallel, z) \nonumber\\
    &= \sum_{\ell \geq 2} \int_{-\Pi_{\rm max}}^{\Pi_{\rm max}} \rmd r_\parallel\, \xi_{\rmg \gamma}^{(\ell)}(r) 
    \mathcal{L}_\ell^2(\mu_{\br}), 
\end{align}
with $\mu_{\br} = \sqrt{r_\parallel^2 / (r_\mathrm{p}^2 + r_\parallel^2)}$.

\section{IA of clusters with varying shape estimators in mock}
\label{append: mock_varying_shape}

We checked how different shape estimators affect the measured IA of galaxy clusters in mock simulation. The shape of galaxy clusters are measured using 
\begin{itemize}
\item dark matter particle distribution (DM),
\begin{equation}
\label{eq:Iij_DM}
    I_{ij}=\frac{\sum_{n} m_n \frac{x_{ni} x_{nj}}{r_n^2}}{\sum_{n} m_n},  
\end{equation}
where $m_n$ is the mass of the $n$th particle within the halo, $x_{ni}$, $x_{nj} (i,j=1,2)$ are the position coordinates of this particle with respect to the centre of cluster, and $r_n$ is the distance of the particle to the cluster center;

\item satellite distribution within dark matter halos (Halo Sat), 

\begin{equation}
\label{eq: Iij_halosat}
    I_{ij}=\frac{\sum_{n} x_{ni} x_{nj}}{N_{g}},  
\end{equation}
where $x_{ni}$, $x_{nj} (i,j=1,2)$ are the positions of $n$th satellite galaxy with respect to the centre of cluster, and $N_{g}$ is the total number of satellite galaxies used for the calculation;

\item $\redmapper$ identified member galaxy distribution (RM Mem), $I_{ij}$ is calculated using Eq. \ref{eq: Iij_halosat}, except that we use member galaxies identified by the $\redmapper$ cluster finder;

\item $\redmapper$ identified $\redmapper$ members that truly belong to the clusters (RM True Mem), also using Eq. \ref{eq: Iij_halosat}.
\end{itemize}

\begin{figure}
    \centering
    \includegraphics[width=.95\linewidth]{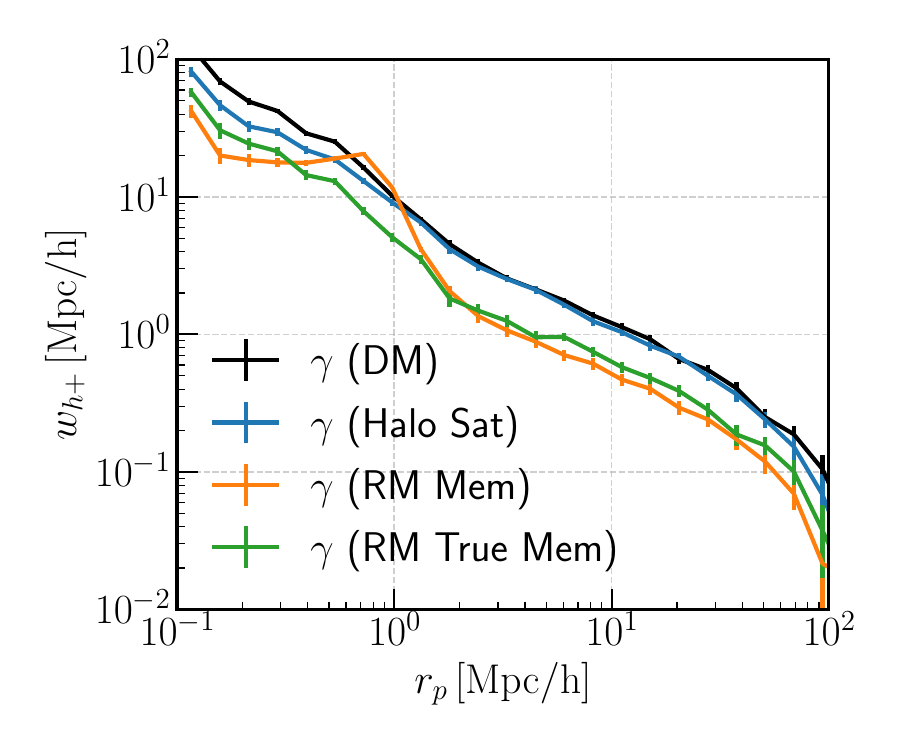}
    \caption{The IA of clusters in the mock clusters ($20\le\lambda_{\rm obs}<200$), with different shape estimators. The black, blue, orange, and green lines show the measurement using shapes estimated using DM particles, satellites within dark matter halos, $\redmapper$ identified cluster members, and true cluster members identified by $\redmapper$. Note that here we show the results for one simulation realization, the results are same across all the realizations. }
    \label{fig:wgp_vary_shape}
\end{figure}

Figure~\ref{fig:wgp_vary_shape} showed that, $w_{h+}$ measured using $\gamma$ (DM) shows the strongest signal, and satellite distributions trace the DM distribution rather well, showing only a slightly weaker IA signal, as shown by the blue line. This is expected since the satellite galaxies are populated following the dark matter distribution.
IA measured using $\redmapper$ identified member galaxy distribution $\gamma$ (RM Mem) show the lowest signal, with a bump at $r_p \sim 0.8\mpc$. If interlopers are removed for the shape calculation, the bump disappears and the IA signal increases a little bit, shown by the green line. However, the IA signal is still much lower than the one measured using DM and satellite galaxy distribution, indicating that another factor, i.e. the satellites that are missed by $\redmapper$ algorithm, is also responsible for decreasing the IA signal.

\section{Clusters of various richness bins in the Mock}
\label{append: mock_lambda_bins}
\begin{figure}
    \centering
    \includegraphics[width=1.\linewidth]{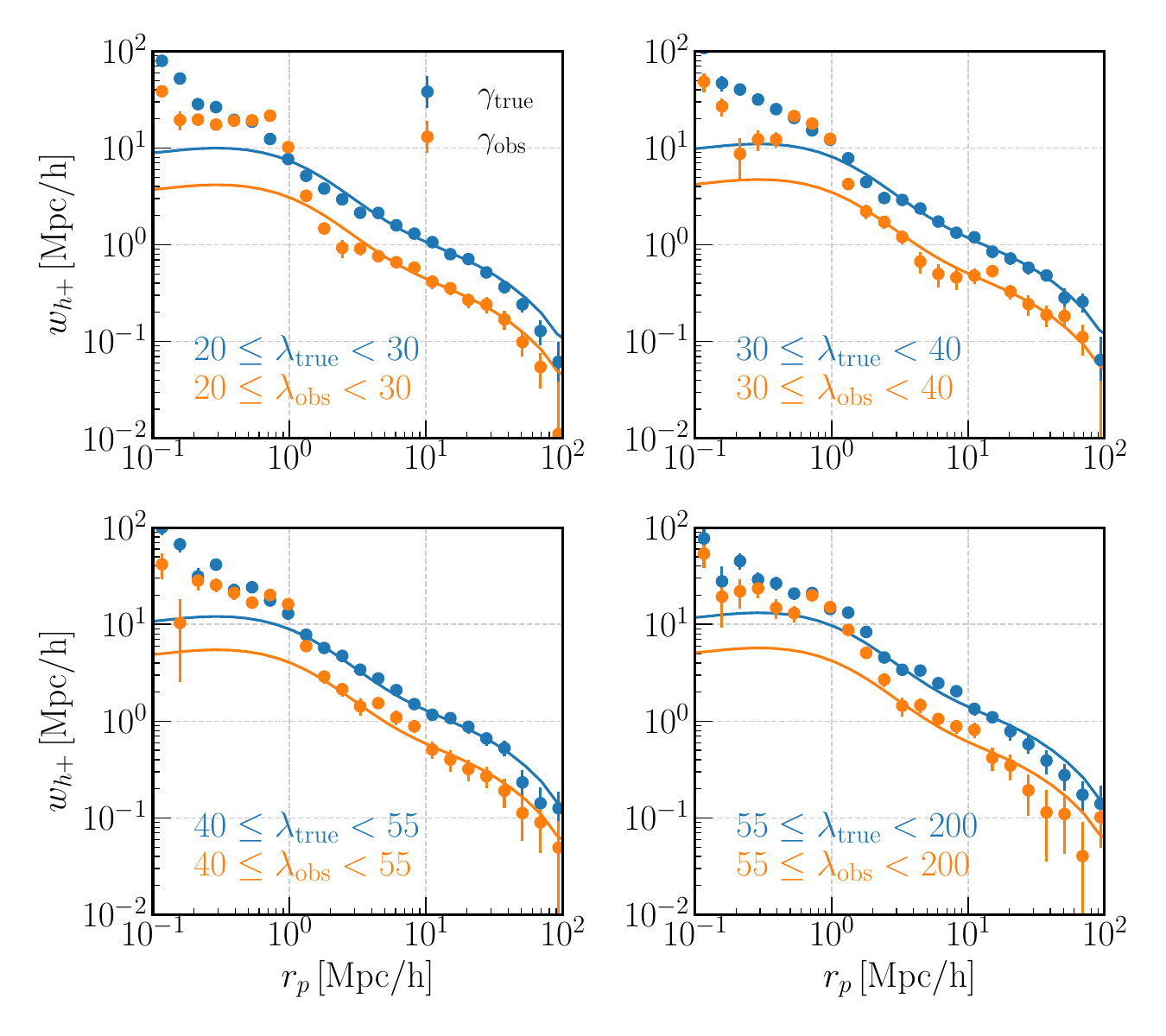}
    \caption{The IA of mock observe (orange) and mock true (blue) clusters, in various richness bins in one simulation realization. The dots are measurement and lines are fitting using NLA in the range of $6\mpc<r_p<70\mpc$. The error bars are estimated using jackknife sub-samples. }
    \label{fig:wgp_mock_lambins}
\end{figure}

Figure~\ref{fig:wgp_mock_lambins} shows the IA of clusters in the mock in various richness bins and the corresponding NLA fitting results. The IA signal of mock true samples are obtained by selecting clusters using $\lambda_{\rm true}$ and measuring shapes $\gamma_{\rm true}$ using satellites within halos. The IA signal of mock observe samples are gotten by selecting clusters using $\lambda_{\rm obs}$ and measuring shapes $\gamma_{\rm obs}$ using $\redmapper$ identified cluster members as in observation. The IA of mock observe is lower than that of mock true in all richness bins. The NLA model fits the signal well in the range of $6\mpc<r_p<70\mpc$, and the resulting $A_{\rm IA}$ are summarized in Table~\ref{tab_cluster}.

\section{Cluster orientation and projection effects}
\label{append: Pmu}

\begin{figure*}
    \centering
    \includegraphics[width=.65\linewidth]{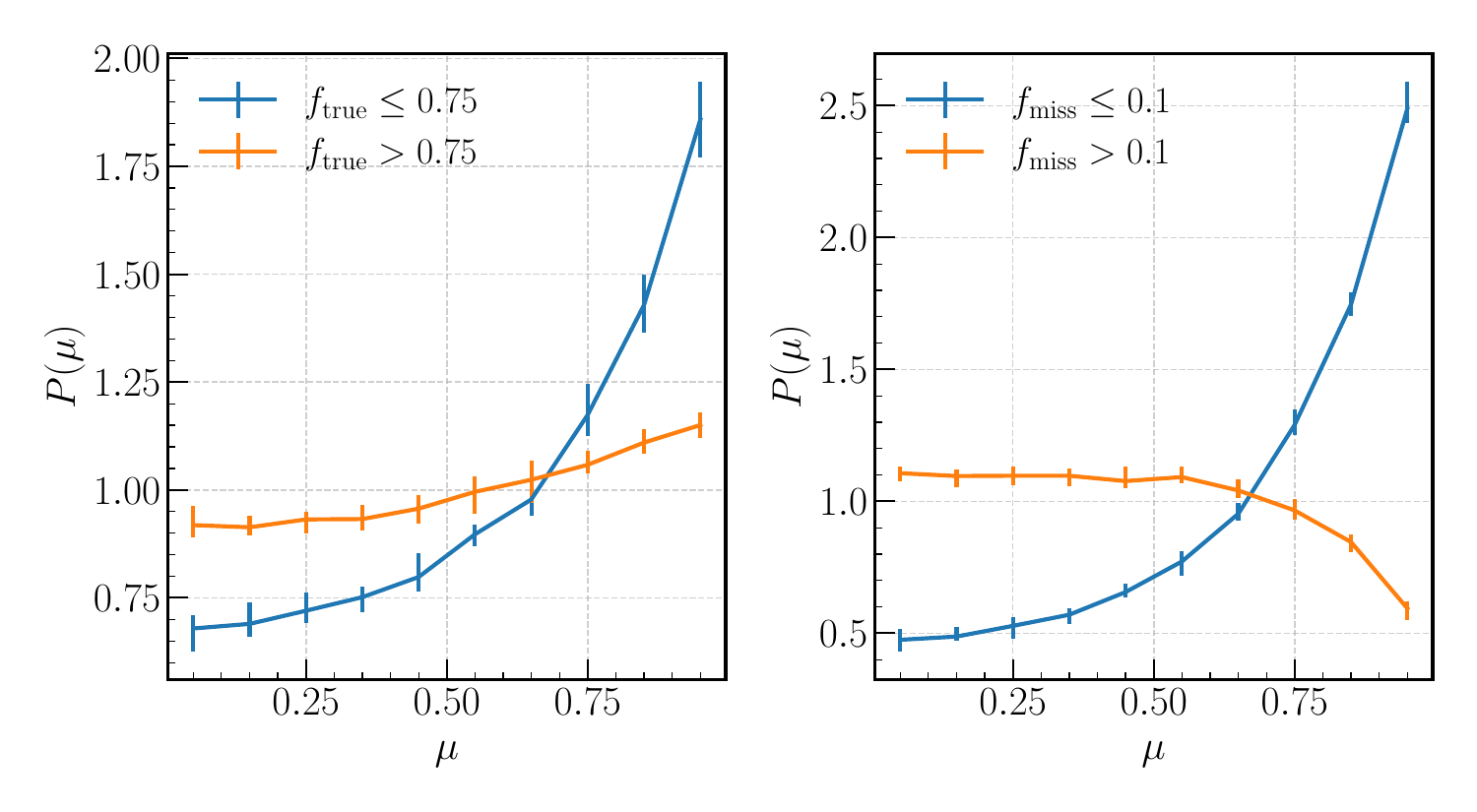}
    \caption{Distribution of cosine of the angle, $\theta$, between major axis of the halo and LOS direction for low/high $f_{\rm true}$ and low/high $f_{\rm miss}$ subsamples with $20\le \lambda_{\rm obs}<200$. Here $\mu\equiv |\cos\theta|$. The lines are median distribution of $19$ realizations, and the error bars show the $1\sigma$ dispersion among the realizations.}
    \label{fig:Pmu}
\end{figure*}

Figure~\ref{fig:Pmu} shows the distribution of the orientation of clusters with respect to LOS direction for clusters with lower and higher $f_{\rm true}$ ($f_{\rm miss}$) separately. The cluster orientation is obtained by calculating the major eigen vectors from the $3$-dimensional inertia tensor using dark matter particle distribution,
\begin{equation}
\label{eq:Iij_DM_3D}
    I_{ij}=\frac{\sum_{n} m_n \frac{x_{ni} x_{nj}}{r_n^2}}{\sum_{n} m_n},  
\end{equation}
where $x_{ni}$, $x_{nj}\ (i,j=1,2,3)$ are the positions of $n$th particle with respect to the centre of cluster. The angle between major axis of the halo and LOS direction is characterized by $\mu\equiv |\cos\theta|$. Clusters selected using $f_{\rm true}\le 0.75$ or $f_{\rm miss}\le 0.1$ tend to have their major axis parallel with the LOS direction. On the other hand, clusters with $f_{\rm true}>0.75$ do not show a strong orientation preference. Cluster with $f_{\rm miss}>0.1$ show a clear tendency of major axis perpendicular to the LOS direction. Figure~\ref{fig:Pmu} shows the distribution for clusters with $20\le \lambda_{\rm obs}<200$ only. The results stay the same when we use different $\lambda_{\rm obs}$ ranges. 

\section{Dependence on halo mass and redshift of $A_{\rm IA}$}
\label{append: aia_mass}

\begin{figure}
    \centering
    \includegraphics[width=.95\linewidth]{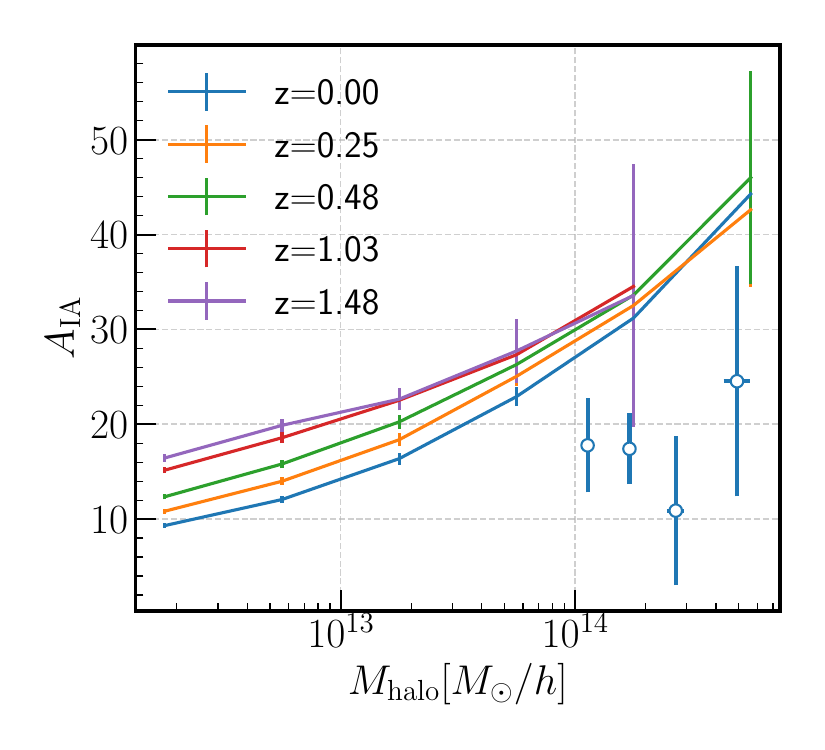}
    \caption{$A_{\rm IA}$ as a function of halo mass and redshift. The lines with error bars are calculated from dark matter halos in N-body simulation as done in \citet{2021MNRAS.501..833K}. The dots are results using SDSS DR8 $\redmapper$ catalog.}
    \label{fig:aia_hmass}
\end{figure}

Figure~\ref{fig:aia_hmass} shows how $A_{\rm IA}$ varies with halo mass and redshift. The lines are results obtained from simulations, where the halo shapes are measured using Eq.~\ref{eq:Iij_DM}, the dots with error bars are results from observation. The halo mass of $\redmapper$ clusters are obtained using the mass - richness relation from \citet{2017MNRAS.466.3103S}, where weak lensing analysis was preformed for the $\redmapper$ clusters at $0.1<z\le 0.33$. \citet{2017MNRAS.466.3103S} Parameterized the relation as $M=M_0 (\lambda/\lambda_0)^{\alpha}$, where ${\rm log}M_0  = 14.344 \pm 0.031$, $\alpha=1.33^{+0.09}_{-0.10}$, and $\lambda_0=40$. We use the mean richness value of each sub-sample to do the conversion. The simulation shows that $A_{\rm IA}$ increases with halo mass and redshift. However, the redshift dependence is very weak/almost gone for halos with $M_h>10^{14}\msun$. The observed $A_{\rm IA}$-$M_{h}$ relation is clearly much lower than that from dark matter halo simulation, which is mainly due to the projection effects. 

\end{document}